\documentclass[aps,twocolumn, noshowpacs, floatfix]{revtex4}
\usepackage{}

\usepackage{amsfonts}
\usepackage{amssymb}
\usepackage{graphicx}
\usepackage{amsmath}
\usepackage[english]{babel}
\usepackage{color}

\begin{document}

\title{Fast, accurate simulation of polaron dynamics and multidimensional spectroscopy by multiple Davydov trial states
}

\author{Nengji Zhou$^{1,2}$, Lipeng Chen$^{2}$, Zhongkai Huang$^{2}$, Kewei Sun$^{3}$,
Yoshitaka Tanimura$^{4}$, and Yang Zhao$^{2}$\footnote{Electronic address:~\url{YZhao@ntu.edu.sg}}}
\date{\today}
\affiliation{$^1$Department of Physics, Hangzhou Normal University, Hangzhou 310046, China \\
$^2$Division of Materials Science, Nanyang Technological University, Singapore 639798, Singapore\\
$^3$School of Science, Hangzhou Dianzi University, Hangzhou 310018, China\\
$^4$Department of Chemistry, Graduate School of Science, Kyoto University, Kyoto 606-8502, Japan
}

\begin{abstract}
By employing the Dirac-Frenkel time-dependent variational principle,
we study the dynamical properties of the Holstein molecular crystal
model with diagonal and off-diagonal exciton-phonon coupling.  A linear
combination of the Davydov D$_1$ (D$_2$) {\it Anstaz}, referred to as the ``multi-D$_1$ {\it
Ansatz}" (``multi-D$_2$ {\it Ansatz}"), is used as the trial state with
enhanced accuracy but without sacrificing efficiency. The time evolution of the
exciton probability is found to be in perfect agreement with that
of the hierarchy equations of motion, demonstrating the
promise the multiple Davydov trial states hold as an efficient,
robust description of dynamics of complex quantum systems. In addition
to the linear absorption spectra computed for both diagonal and
off-diagonal cases, for the first time, $2$D spectra have been
calculated for systems with off-diagonal exciton-phonon coupling by
employing the multiple ${\rm D}_2$ {\it Ansatz} to compute the
nonlinear response function, testifying to the great potential of the multiple ${\rm D}_2$ {\it Ansatz}
for fast, accurate implementation of multidimensional
spectroscopy. It is found that the signal exhibits a
single peak for weak off-diagonal coupling, while a vibronic
multi-peak structure appears for strong off-diagonal coupling.

\end{abstract}

\maketitle

\section{Introduction}

Thanks to recent advances in ultrafast spectroscopy, femtosecond
photoexcitation has became a major technique in probing elementary
excitations, which brought about numerous studies on relaxation
dynamics of photoexcited entities, for example, polarons in
inorganic liquids and solids \cite{an_04, zheng_06, liu_06}, charge
carriers in topological insulators \cite{bron_14, tim_12}, trapped
electrons and holes in the semiconductor nanoparticles \cite{chi_02,
kli_00,kim_96}, and electron-hole pairs in light-harvesting
complexes of photosynthesis \cite{sau_79, reng_01, blan_02,
gron_06,LP_Molecule}. Emerging technological capabilities to control
femtosecond pulse durations and down-to-one-hertz bandwidth
resolutions offer unpreceded windows on vibrational dynamics and
excitation relaxation. For example, progress in femtosecond
spectroscopy has enabled the observation of a coherent phonon wave
packet oscillating along an adiabatic potential surface associated
with a self-trapped exciton in a crystal with strong exciton-phonon
interactions \cite{tom_00}. Taking advantage of the ultrashort pulse
widths of recent lasers, the femtosecond dynamics of polaron
formation and exciton-phonon dressing have been observed in
pump-probe experiments \cite{dex_00, sug_01, mor_10}. These
experiments have revealed a complex interplay between a single
exciton and its surrounding phonons under nonequilibrium conditions,
while theoretical developments have not been kept in parallel. In
particular, modeling of polaron dynamics have not received
much-deserved attention over the last six decades \cite{ale_95,
pee_84}.

\begin{figure}[tbp]
\centering
\includegraphics[scale=0.3]{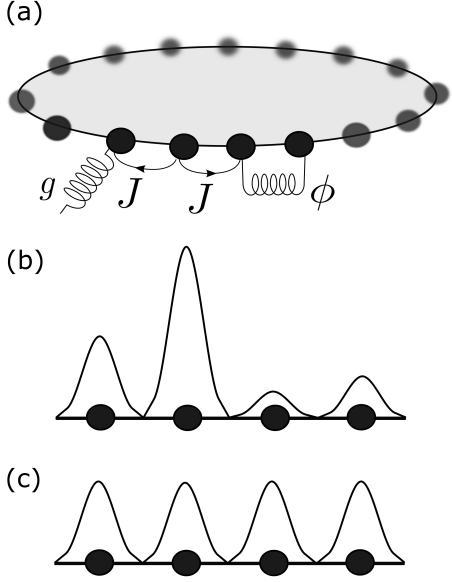}
\caption{(a) Schematic of the Holstein ring. A simplified molecular crystal is treated as a ring where each point represents a big molecule and wave lines inside denote the phonons. (b) and (c) Schematics of waveshapes for the Davydov $\rm D_1$ and $\rm D_2$ {\it Ans\"atze}, respectively. The phonon part of the $\rm D_1$ {\it Ansatz} depends on both sites and momentum, while that of the $\rm D_2$ {\it Ansatz} is site independent.}
\label{fig1_ring}
\end{figure}

From a theoretical point of view, capturing time-dependent polaron
formation requires an in-depth understanding of the combined
dynamics of the particle and the phonons in its environment
\cite{ran_06}. A simple Hamiltonian is that of the extended Holstein
molecular crystal model \cite{hol_59, hol_59_2} with simultaneous
diagonal and off-diagonal exciton-phonon coupling, as shown in
Fig.~\ref{fig1_ring}(a), where the diagonal coupling represents a
nontrivial dependence of the exciton site energies on the lattice coordinates,
and the off-diagonal coupling, a nontrivial dependence of the exciton transfer integral
on the lattice coordinates \cite{su_79,dis_84,
sumi_89,zhao_94,dmchen_11}. A large body of literature exists on the
study of the conventional form of the Holstein Hamiltonian with the
diagonal coupling only \cite{luo_10, sun_10}. It seems fundamental
to take into account simultaneously diagonal and off-diagonal
coupling to characterize solid-state excimers \cite{dis_84, sumi_89}
as a variety of experimental and theoretical studies imply a strong
dependence of electronic tunneling upon certain coordinated
distortions of neighboring molecules in the formation of bound
excited states. However, complete understanding of the off-diagonal
coupling and out-of-equilibrium phenomena remains elusive. Early
treatments of the off-diagonal coupling include the Munn-Silbey
theory \cite{MunnSilbey,zhao_94, dmchen_11}, which is based upon a perturbative
approach with additional constraints on canonical transformation
coefficients determined by a self-consistency equation. The
global-local (GL){\it Ansatz} \cite{zhao_94b, zhao_08}, formulated
by Zhao and co-workers in the early $1990$s, was subsequently
employed in combination with the dynamic coherent potential
approximation (with the Hartree approximation) to arrive at a
state-of-the-art ground-state wave function as well as higher
eigenstates \cite{liu_09}.

Because an exact solution of the polaron dynamic still eludes us,
several numerical approaches have been developed. For example, the
time-dependent Schr\"odinger equation can be numerically
integrated in real space for a few phonon time periods to probe the
time evolution of electron and phonon densities and electron-phonon
correlation functions \cite{ku_07}. However, the method is time
consuming and impractical when the size of the system is large.
Fortunately, time-dependent variational approaches are still valid
to treat the polaron dynamics in such cases as long as a proper
trial wave function is adopted. Previously, static properties of the
Holstein polaron were studied by Zhao and his co-workers with a set
of trial wave functions based upon phonon coherent states, including
the Toyozawa {\it Ansatz} \cite{zhao_94b, meier_97,zhao_97}, the GL
{\it Ansatz} \cite{zhao_94b, zhao_97, zhao_08, zhao_97b}, a
delocalized form of the Davydov $\rm D_1$ {\it Ansatz} \cite{sun_13},
and the multi-$\rm D_1$ {\it Ansatz} \cite{zhou_14}. The results of
these extended Davydov {\it Ans\"atze} exhibit great promises in the
investigation of the polaron energy band and other static properties
of the Holstein polaron. However, difficulties surround accurate
simulations of the polaron dynamics from an arbitrary initial state,
such as a localized state for which the aforementioned Bloch states
are not well suited. Thus, the question of what type of the
variational trial state is suitable for the polaron dynamics of the
Holstein model is still open.

By using the Dirac-Frenkel time-dependent variational principle, a
powerful apparatus to reveal accurate dynamics of quantum many-body
systems \cite{dira_30}, one can study the polaron dynamics of the
Holstein model with the simultaneous diagonal and off-diagonal
exciton-phonon coupling. Time-dependent variational parameters,
which specify the trial state, are obtained by solving a set of
coupled differential equations generated from the Lagrangian
formalism of the Dirac-Frenkel variation. Validity of the trial
states is carefully examined by quantifying how faithfully they
follow the Schr\"odinger equation \cite{luo_10, sun_10, zhao_12}.
The hierarchy of the Davydov {\it Ans\"atze} includes two trial
states of varying sophistication, referred to as the $\rm D_1$ and
$\rm D_2$ {\it Ans\"atze} \cite{Davydov1,Davydov2,skri_88, for_93, han_94}, with the
latter being a simplified version of the former. The ${\rm D_1}$
{\it Ansatz} is sufficient to describe the Holstein polaron dynamics
with the diagonal coupling, but fails in the presence of the
off-diagonal coupling. In comparison, the ${\rm D_2}$ {\it Ansatz}
exhibits a nice dynamical performance with the off-diagonal
coupling, though the deviation from the exact solution to the
Schr\"odinger dynamics is not disregarded \cite{zhao_12}. Instead,
superposition of the ${\rm D_1}$ or the ${\rm D_2}$ {\it Ansatz}
will be adopted in our work, which offers significant improvements
in the flexibility of the trial state \cite{zhou_15}, thus yielding
accurate polaron dynamics of the Holstein model with the
simultaneous diagonal and off-diagonal coupling.

Recently, two dimensional (2D) electronic spectroscopy has been widely used to probe ultrafast energy transfer and charge separation processes in photosynthetic light harvesting complexes \cite{Brixner,Engel,Collini,Panitchayangkoon,Myers,Lewis,Romero}. Compared to linear spectroscopy techniques in which the spectral lines are often congested, ultrafast non-linear spectroscopies can resolve dynamical processes with various time scales. In a 2D electronic spectroscopy experiment and apparatus, for example, three ultra-short laser pulses, separated by two time delays, namely, the coherence time and the waiting time, are incident on the sample, and the resultant signal field is spectrally resolved in a given phase-matched direction. The 2D contour plots of the signals provide direct information about excitonic relaxation and dephasing in a variety of molecular systems. Simulation of 2D electronic spectra of molecular aggregates was previously carried out for the Holstein model with diagonal exciton-phonon coupling. However, the effect of off-diagonal coupling on the 2D spectra is yet to be addressed.

In this paper, the multiple Davydov trial states, called the
multi-$\rm D_1$ and multi-$\rm D_2$ {\it Ans\"atze}, will be adopted
to simulate the polaron dynamics of an extended Holstein Hamiltonian
that includes the off-diagonal exciton-phonon coupling. Validity of
these trial states is carefully examined with the linear absorption
spectra compared closely with the ground-state energy band. In addition, 2D spectra for systems with off-diagonal exciton phonon coupling will be calculated by employing the multiple $\mathrm{D}_2$ {\it Ansatz}. The
remainder of the paper is organized as follows. In Sec.~II we
introduce the Holstein Hamiltonian and two novel variational wave
functions on the basis of the multiple Davydov trial states,
together with a criterion that quantifies the deviation of our trial
states from the solution to the Schr\"odinger equation. In Sec.~III,
results are analyzed including the time evolution of the exciton
amplitudes and the phonon displacements, the quantitative
measurement for the trial state validity, and the linear absorption
and 2D spectra. Finally, conclusions are drawn in Sec.~IV.

\section{METHODOLOGY}

\subsection{Model}


The Hamiltonian of the one-dimensional Holstein polaron is composed of
\begin{equation}
\hat{H}=\hat{H}_{\rm ex}+\hat{H}_{\rm ph}+\hat{H}_{\rm ex-ph}^{\rm diag}+\hat{H}_{\rm ex-ph}^{\rm o.d.}
\label{Holstein}
\end{equation}
where $\hat{H}_{\rm ex},\hat{H}_{\rm ph},\hat{H}_{\rm ex-ph}^{\rm diag}$ and $\hat{H}_{\rm ex-ph}^{\rm o.d.}$ represent
the exciton Hamiltonian, bath (phonon) Hamiltonian, diagonal exciton-phonon
coupling Hamiltonian and off-diagonal coupling Hamiltonian, respectively, which are defined as
\begin{eqnarray}
\hat{H}_{\rm ex} & = & -J\sum_{n}\hat{a}_{n}^{\dagger}\left(\hat{a}_{n+1}+\hat{a}_{n-1}\right), \nonumber \\
\hat{H}_{\rm ph} & = & \sum_{q}\omega_{q}\hat{b}_{q}^{\dagger}\hat{b}_{q},   \\
\hat{H}_{\rm ex-ph}^{\rm diag} & = & -g\sum_{n,q}\omega_{q}\hat{a}_{n}^{\dagger}\hat{a}_{n}\left(e^{iqn}\hat{b}_{q}+e^{-iqn}\hat{b}_{q}^{\dagger}\right), \nonumber \\
\hat{H}_{\rm ex-ph}^{\rm o.d.} & = & \frac{1}{2}\phi\sum_{n,q}\omega_{q}\{\hat{a}_{n}^{\dagger}\hat{a}_{n+1}[e^{iqn}(e^{iq}-1)\hat{b}_{q}+ {\rm H.c.}], \nonumber \\
 &  & +\hat{a}_{n}^{\dagger}\hat{a}_{n-1}[e^{iqn}(1-e^{-iq})\hat{b}_{q}+ {\rm H.c.}]\}, \nonumber
\end{eqnarray}
where $\rm H.c.$ denotes the Hermitian conjugate, $\omega_q$ is the phonon frequency with momentum $q$, $\hat{a}_n^{\dag}$ ($\hat{a}_n$) is the exciton creation (annihilation) operator
for the $n$-th molecule, and $\hat{b}_q^{\dag}$ ($\hat{b}_q$) is the creation (annihilation) operator of a phonon with the momentum $q$,
\begin{equation}
\hat{b}_q^{\dag} = N^{-1/2}\sum_n e^{iqn}\hat{b}_n^{\dag}, \quad \hat{b}_n^{\dag} =  N^{-1/2}\sum_q e^{-iqn}\hat{b}_q^{\dag}.
\label{momentum}
\end{equation}
The parameters $J, g$ and $\phi$ represent the transfer integral, diagonal and off-diagonal coupling strengthes, respectively, and $N=16$ is the number of sites in the Holstein ring.
In this paper, a linear phonon dispersion is assumed,
\begin{equation}
\omega_{q}=\omega_{0}\left[1+W(\frac{2\left|q\right|}{\pi}-1)\right],
\end{equation}
where $\omega_0$ denotes a central phonon frequency, $W$ is the band width falling between $0$ and $1$, and
$q=2\pi l/N$ represents the momentum index with $l=-\frac{N}{2}+1, \ldots, \frac{N}{2}$.

\subsection{Multiple Davydov trial states}

In the past, two typical Davydov trial states, i.e., the $\rm D_1$ and $\rm D_2$
{\it Ans\"atze}, were used to obtain the time evolution of the Holstein polaron
following the Dirac-Frenkel variation scheme. The $\rm D_2$ {\it Ansatz} is a simplified version of the $\rm D_1$ {\it Ansatz},
since the phonon displacements of the $\rm D_1$ ($\rm D_2$) trial state is site-dependent (site-independent), as illustrated in Figs.~\ref{fig1_ring}(b) and (c).
Multiple Davydov trial state with the multiplicity $M$ are then introduced in this paper, which can be constructed as follows
\begin{eqnarray} \label{D1_state}
&& \left|{\rm D_1^M}\left(t\right)\right\rangle  =  \sum_{i}^{M}\sum_{n}^N\psi_{i,n}\left|n\right\rangle \left|\lambda_{i,n}\right\rangle,  \\ \nonumber
&& =\sum_{i}^{M}\sum_{n}^N\psi_{i,n} \hat{a}_{n}^{\dagger}\left|0\right\rangle _{\rm ex} \exp\left\{ \sum_{q}\left[\lambda_{inq}\hat{b}_{q}^{\dagger}-\lambda_{inq}^{\ast}\hat{b}_{q}\right]\right\} \left|0\right\rangle _{\rm ph},
\end{eqnarray}
and
\begin{eqnarray}\label{D2_state}
&& \left|{\rm D_2^M}\left(t\right)\right\rangle  =  \sum_{i}^{M}\sum_{n}^N\psi_{i,n}\left|n\right\rangle \left|\lambda_{i}\right\rangle, \\ \nonumber
&& =\sum_{i}^{M}\sum_{n}^N\psi_{i,n} \hat{a}_{n}^{\dagger}\left|0\right\rangle _{\rm ex} \exp\left\{ \sum_{q}\left[\lambda_{iq}\hat{b}_{q}^{\dagger}-\lambda_{iq}^{\ast}\hat{b}_{q}\right]\right\} \left|0\right\rangle _{\rm ph},
\end{eqnarray}
where $\psi_{i,n}$ and $\lambda_{inq}$ are related to the exciton probability and the phonon displacement, respectively, $n$ represents the site number in the
molecular ring, and $i$ labels the coherent superposition state. If $M=1$, both the $|\rm D_1^M(t)\rangle$ and $|\rm D_2^M(t)\rangle$ {\it Ans\"atze}
are restored to the usual Davydov $\rm D_1$ and $\rm D_2$ trial states, respectively. The equation of motion of the variational
parameters $\psi_{i,n}$ and $\lambda_{inq}$ are then derived by adopting the Dirac-Frenkel variational principle,

\begin{eqnarray}\label{eq:eom1}
\frac{d}{dt}\left(\frac{\partial L}{\partial\dot{\psi_{i,n}^{\ast}}}\right)-\frac{\partial L}{\partial\psi_{i,n}^{\ast}} & = & 0, \nonumber \\
\frac{d}{dt}\left(\frac{\partial L}{\partial\dot{\lambda_{inq}^{\ast}}}\right)-\frac{\partial L}{\partial\lambda_{inq}^{\ast}} & = & 0.
\end{eqnarray}

For the multi-$\rm D_1$ {\it Ansatz} defined in Eq.~(\ref{D1_state}), the Lagrangian $L_1$ is given as
\begin{eqnarray}
L_{1} & = & \left\langle {\rm D}^M_{1}\left(t\right)\right|\frac{i\hslash}{2}\frac{\overleftrightarrow{\partial}}{\partial t}-\hat{H}\left|{\rm D}^M_{1}\left(t\right)\right\rangle \nonumber \\
& = & \frac{i\hslash}{2}\left[\left\langle {\rm D}^M_{1}\left(t\right)\right|\frac{\overrightarrow{\partial}}{\partial t}\left|{\rm D}^M_{1}\left(t\right)\right\rangle -\left\langle {\rm D}_{1}\left(t\right)\right|\frac{\overleftarrow{\partial}}{\partial t}\left|{\rm D}^M_{1}\left(t\right)\right\rangle \right]\nonumber \\
&  & -\left\langle {\rm D}^M_{1}\left(t\right)\right|\hat{H}\left|{\rm D}^M_{1}\left(t\right)\right\rangle,
\label{Lagrangian_1}
\end{eqnarray}
where the first term yields
\begin{eqnarray}
 &  & \left\langle {\rm D}_1^M\left(t\right)\right|\frac{\overrightarrow{\partial}}{\partial t}\left|{\rm D}_1^M\left(t\right)\right\rangle -\left\langle {\rm D}_1^M\left(t\right)\right|\frac{\overleftarrow{\partial}}{\partial t}\left|{\rm D}_1^M\left(t\right)\right\rangle \nonumber \\
 &  &= \sum_{i,j}^{M}\sum_{n}\left(\psi_{jn}^{\ast}\dot{\psi}_{in}-\dot{\psi}_{jn}^{\ast}\psi_{in}\right)S_{ji}\nonumber \\
 &  & +\sum_{i,j}^{M}\sum_{n}\psi_{jn}^{\ast}\psi_{in}S_{ji}\sum_{q}\left[\frac{\dot{\lambda}_{jnq}^{\ast}\lambda_{jnq}+\lambda_{jnq}^{\ast}\dot{\lambda}_{jnq}}{2}\right.\nonumber \\
 &  & \left.-\frac{\dot{\lambda}_{inq}\lambda_{inq}^{\ast}+\lambda_{inq}\dot{\lambda}_{inq}^{\ast}}{2}+\lambda_{jnq}^{\ast}\dot{\lambda}_{inq}-\lambda_{inq}\dot{\lambda}_{jnq}^{\ast}\right],
\label{energies}
\end{eqnarray}
and the second term is
\begin{eqnarray}
 &  & \left\langle {\rm D}^M_{1}\left(t\right)\right|\hat{H}\left|{\rm D}^M_{1}\left(t\right)\right\rangle \nonumber \\
 &  &= \left\langle {\rm D}^M_{1}\left(t\right)\right|\hat{H}_{\rm ex}\left|{\rm D}^M_{1}\left(t\right)\right\rangle +\left\langle {\rm D}^M_{1}\left(t\right)\right|\hat{H}_{\rm ph}\left|{\rm D}^M_{1}\left(t\right)\right\rangle \nonumber \\
 &  & +\left\langle {\rm D}^M_{1}\left(t\right)\right|\hat{H}_{\rm ex-ph}^{\rm diag}\left|{\rm D}^M_{1}\left(t\right)\right\rangle +\left\langle {\rm D}^M_{1}\left(t\right)\right|\hat{H}_{\rm ex-ph}^{\rm o.d.}\left|{\rm D}^M_{1}\left(t\right)\right\rangle. \nonumber \\
\end{eqnarray}
Detailed derivations on the equations of motion for the variational parameters are given in Appendix A.

Similarly, the equations of motion for the multi-${\rm D}_2$ {\it Ansatz} can be derived using the Dirac-Frenkel variational principle in Eq.~(\ref{eq:eom1}) with the Lagrangian
$L_2$ defined as
\begin{eqnarray}
L_2 & = & \langle {\rm D}^M_2(t)|\frac{i\hbar}{2}\frac{\overleftrightarrow{\partial}}{\partial t}- \hat{H}|{\rm D}^M_2(t)\rangle \nonumber \\
& = & \frac{i\hbar}{2}\left[ \langle {\rm D}^M_2(t)|\frac{\overrightarrow{\partial}}{\partial t}|{\rm D}^M_2(t)\rangle - \langle {\rm D}^M_2(t)|\frac{\overleftarrow{\partial}}{\partial t}|{\rm D}^M_2(t)\rangle \right] \nonumber \\
&-& \langle {\rm D}^M_2(t)|\hat{H}|{\rm D}^M_2(t)\rangle.
\label{Lagrangian_2}
\end{eqnarray}

Assuming the trial wave function $|{\rm D}^M_{1,2}(t)\rangle=|\Psi(t)\rangle$ at the time $t$, we introduce a deviation vector $\vec{\delta}(t)$ to quantify the accuracy of the variational dynamics based on the multiple Davydov trial states,
\begin{eqnarray}
\vec{\delta}(t) & =  & \vec{\chi}(t) -\vec{\gamma}(t)  \nonumber \\
& = & \frac{\partial}{\partial t}|\Psi(t)\rangle - \frac{\partial}{\partial t}|{\rm D}^M_{1,2}(t)\rangle.
\label{deviation_1}
\end{eqnarray}
where the vectors $\vec{\chi}(t)$ and $\vec{\gamma}(t)$ obey the Schr\"{o}dinger equation  $\vec{\chi}(t)=\partial |\Psi(t)\rangle / \partial t = \frac{1}{i\hbar}\hat{H}|\Psi(t)\rangle$ and the Dirac-Frenkel variational dynamics $\vec{\gamma}(t)=\partial |{\rm D}^M_{1,2}\rangle / \partial t$ in Eq.~(\ref{eq:eom1}), respectively. Using the Schr\"{o}dinger equation and the relationship $|\Psi(t)\rangle=|{\rm D}^M_{1,2}(t)\rangle$ at the moment $t$, the deviation vector $\vec{\delta}(t)$ can be calculated as
\begin{equation}
\vec{\delta}(t) = \frac{1}{i\hbar}\hat{H}|{\rm D}^M_{1,2}(t)\rangle - \frac{\partial}{\partial t}|{\rm D}^M_{1,2}(t)\rangle.
\label{deviation_2}
\end{equation}
Thus, deviation from the exact Schr{\"o}dinger dynamics can be indicated by the amplitude of the deviation vector $\Delta(t)=||\vec{\delta}(t)||$.
In order to view the deviation in the parameter space $(W,J,g,\phi)$, a dimensionless relative deviation $\sigma$ is calculated as
\begin{equation}
\sigma = \frac{{\rm max}\{\Delta(t)\} }{{\rm mean}\{N_{\rm err}(t)\}}, \quad \quad t \in [0, t_{\rm max}].
\label{relative_error}
\end{equation}
where $N_{\rm err}(t)=||\vec{\chi}(t)||$ is the amplitude of the time derivative of the wave function,
\begin{eqnarray}
N_{\rm err}(t) & = & \sqrt{\langle\frac{\partial}{\partial t}\Psi(t)|\frac{\partial}{\partial t}\Psi(t)\rangle} \nonumber \\
& = & \sqrt{\langle {\rm D}^M_{1,2}(t)|\hat{H}^2|{\rm D}^M_{1,2}(t)\rangle} \nonumber \\
& \approx & \Delta E,
\end{eqnarray}
since $\langle E \rangle = \langle {\rm D}^M_2(t)|\hat{H}(t)|{\rm D}^M_2(t)\rangle \approx 0$ in this paper.

Two types of initial states are considered, i.e.,
the exciton of the Holstein polaron either sits on a single site for diagonal coupling cases or on two nearest-neighboring sites for off-diagonal coupling cases.
Other initial states, such as Gaussian distributed and uniformly occupied, have also been investigated,
leading to similar results but with larger relative errors. To avoid singularity, noise satisfying the uniform distribution $[-10^{-5}, 10^{-5}]$
is added to the variational parameters $\psi_{i,n}$ and $\lambda_{iq}$ ($\lambda_{inq}$) of the initial states.
With the wave functions $|{\rm D}^M_1(t)\rangle$ and $|{\rm D}^M_2(t)\rangle$ at hand, the energy of the Holstein polaron
$E_{\rm total} = E_{\rm ex}+ E_{\rm ph}+ E_{\rm diag}+ E_{\rm off}$ is calculated, where
$E_{\rm ex} = \langle {\rm D}^M_{1,2}|\hat{H}_{\rm ex}|{\rm D}^M_{1,2} \rangle, ~ E_{\rm ph} = \langle {\rm D}^M_{1,2}|\hat{H}_{\rm ph}|{\rm D}^M_{1,2} \rangle, ~ E_{\rm diag} = \langle {\rm D}^M_{1,2}|\hat{H}_{\rm ex-ph}^{\rm diag}|{\rm D}^M_{1,2} \rangle$ and $E_{\rm off} = \langle {\rm D}^M_{1,2}|\hat{H}_{\rm ex-ph}^{\rm o.d.}|{\rm D}^M_{1,2} \rangle$. In addition, the exciton probability $P_{\rm ex}(t, n)$ and the phonon displacement $X_{\rm ph}(t, n)$ are also calculated
\begin{eqnarray}
P_{\rm ex}(t, n) = \langle {\rm D}^M_{1,2} |\hat{a}_{n}^\dag\hat{a}_{n}|{\rm D}^M_{1,2}\rangle, \nonumber \\
X_{\rm ph}(t, n) = \langle {\rm D}^M_{1,2} |\hat{b}_{n}+\hat{b}_{n}^\dag|{\rm D}^M_{1,2}\rangle.
\label{exciton-phonon}
\end{eqnarray}

\begin{figure}[tbp]
\centering
\includegraphics[scale=0.35]{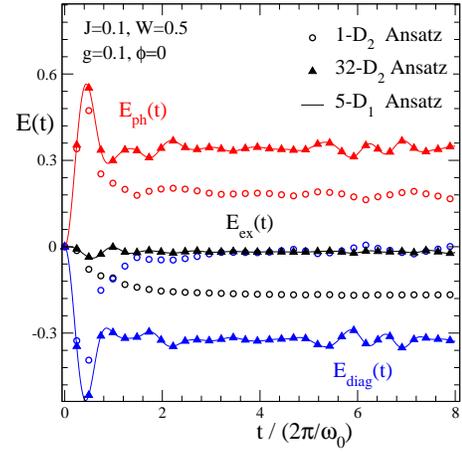}
\caption{The energies of the exciton, phonon, and exciton-phonon interaction, i.e., $E_{\rm ex}(t), E_{\rm ph}(t)$ and $E_{\rm diag}(t)$, are displayed as a function
of the time $t$ for the weak coupling case of $J=0.1, g=0.1, W=0.5$ and $\phi=0$. The open circles, solid triangles and solid line correspond to the results obtained with
the single $\rm D_2$, ${\rm D}_2^{M=32}$ and ${\rm D}_1^{M=5}$ {\it Ans\"atze}, respectively.  }
\label{compare}
\end{figure}

Optical spectroscopy is another important aspect for the
investigation of the polaron dynamics, as it provides valuable
information on various correlation functions.  First of all, the
linear absorption spectra $F(\omega)$ calculated from the polaron
dynamics on the basis of different {\it Ans\"atze} will be
comprehensively studied. The autocorrelation function
$F(t)$ of the exciton-phonon system is introduced
\begin{eqnarray}
F(t) &=& _{\rm ph}\langle0|_{\rm ex}\langle0| e^{i\hat{H}t}\hat{P}e^{-i\hat{H}t}\hat{P}^\dag |0\rangle_{\rm ex}|0\rangle_{\rm ph} \nonumber \\
& = & _{\rm ph}\langle0|_{\rm ex}\langle0| \hat{P}e^{-i\hat{H}t}\hat{P}^\dag |0\rangle_{\rm ex}|0\rangle_{\rm ph},
\label{auto_cor}
\end{eqnarray}
with the polarization operator
\begin{equation}
\hat{P} = \mu \sum_n(\hat{a}_n^{\dag}|0\rangle_{\rm ex}~_{\rm ex}\langle0|+|0\rangle_{\rm ex}~_{\rm ex}\langle0|\hat{a}_n).
\label{polarization}
\end{equation}
The linear absorption $F(\omega)$ is then calculated by means of the Fourier transformation,
\begin{equation}
F(\omega) = \frac{1}{\pi}{\rm Re}\int_0^{\infty} F(t)e^{i\omega t}dt.
\label{linear_function}
\end{equation}
In addition to the information provided by the linear absorption spectra, $2$D electronic spectra provide direct knowledge on exciton-exciton interactions and dephasing and relaxation processes that are elusive in the output from the traditional 1D spectroscopy. Theoretical simulation of 2D spectra involves the calculation of third order polarization $P(t)$, which can be expressed in terms of the nonlinear response function $R_i$, where $i$ goes from 1 to 4 \cite{Mukamel,SunKeWei1,SunKeWei2}. The 2D electronic spectra are measured in two configurations that correspond to the rephasing (subscript R) and non-rephasing (subscript NR) contribution to the third order polarization $P(t)$, which, in the impulsive approximation, can be written as
\begin{eqnarray}\label{Response}
P_{R}^{(3)}(t,T,\tau)\sim{-i}[R_2(t,T,\tau)+R_3(t,T,\tau)], \nonumber\\
P_{NR}^{(3)}(t,T,\tau)\sim{-i}[R_1(t,T,\tau)+R_4(t,T,\tau)].
\end{eqnarray}
Where $\tau$ (the so-called coherence time) is the delay time between the first and second pulses, $T$ (the so-called population time) is the delay time between the second and third pulses, and $t$ is the delay time between the third pulse and measured signal. The rephasing and non-rephasing 2D spectra can be then obtained by performing two-dimensional Fourier-Laplace transformation of Eq.~(\ref{Response}) as follows
\begin{eqnarray}
S_R(\omega_t,T,\omega_{\tau})=\mathrm{Re}\int_0^{\infty}\int_0^{\infty}dtd\tau{i}P_R^{(3)}(t,T,\tau)e^{-i\omega_{\tau}\tau+i\omega_t{t}}, \nonumber
\\
S_{NR}(\omega_t,T,\omega_{\tau})=\mathrm{Re}\int_0^{\infty}\int_0^{\infty}dtd\tau{i}P_{NR}^{(3)}(t,T,\tau)e^{i\omega_{\tau}\tau+i\omega_t{t}}. \nonumber
\\
\end{eqnarray}
The total $2$D signal is defined as the sum of the non-rephasing and the rephasing part
\begin{equation}
S(\omega_t,T,\omega_{\tau})=S_R(\omega_t,T,\omega_{\tau})+S_{NR}(\omega_t,T,\omega_{\tau}).
\end{equation}
In this work, we will apply the multiple ${\rm D}_2$ states to calculate the nonlinear response functions $R_i$ with special attention paid to the role of off-diagonal exciton-phonon coupling on the 2D spectra. The reader is referred to the Appendix D for more details on the applications of the multiple ${\rm D}_2$ {\it Ansatze} to the simulation of $2$D spectra.

\begin{figure}[tbp]
\centering
\includegraphics[scale=0.55]{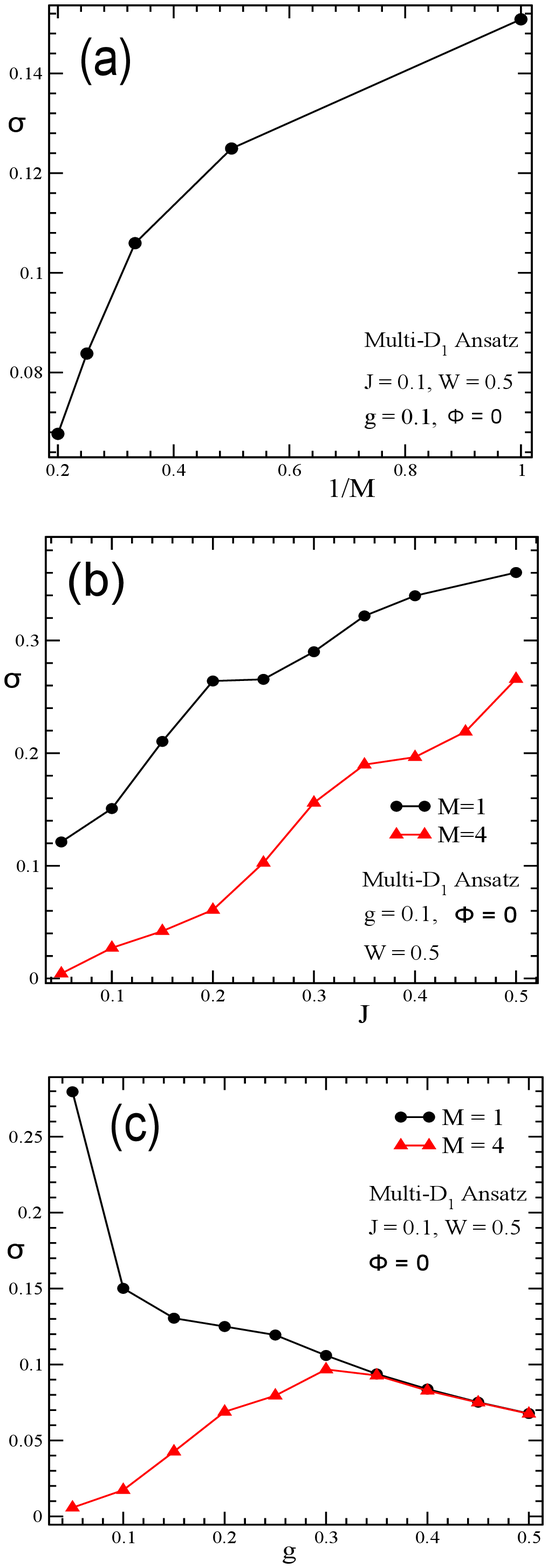}
\caption{(a) The relative deviation $\sigma$ of the multi-${\rm D_1}$ {\it Ansatz} in a $16$-site molecular ring is displayed as a function of $1/M$ representing the inverse of the multiplicity. The set of parameters $J=0.1, g=1, W=0.5$ and $\phi = 0$ is used. Moreover, the relative deviation $\sigma$ for the diagonal coupling case is also plotted as a function
of the transfer integral $J$ in (b) and diagonal coupling strength $g$ in (c). In both of them, the lines with circles and triangles correspond to the results obtained at the multiplicity $M=1$ and $M=4$, respectively.  }
\label{g=0.1}
\end{figure}

\section{Numerical results}
\subsection{Validity of variational dynamics}

Figure~\ref{compare} illustrates the time evolution of the system
energies, including the exciton energy $E_{\rm ex}$, the phonon
energy $E_{\rm ph}$ and the exciton-phonon interaction energy
$E_{\rm diag}$, in a diagonal coupling only case with transfer
integral $J=0.1$, band width $W=0.5$ and coupling strength $g=0.1$.
For a molecular ring of $N=16$ sites, the energies obtained with
three different {\it Ans\"atze} are compared (the open circles
corresponding to the single $\rm D_2$ {\it Ansatz}, the solid
triangles corresponding to the $\rm D_2^{M=32}$ {\it Ansatz}, and
the solid line corresponding to the $\rm D_1^{M=5}$ {\it Ansatz}).
Results obtained with the multi-$\rm D_2$ {\it Ansatz} with $M=32$
display obvious deviations from those by the single $\rm D_2$ {\it Ansatz},
demonstrating the improvement produced by the multiple Davydov trial states
over its single {\it Ansatz} counterpart. In addition, the dynamics generated on the $\rm
D_1$ trial state can be made more accurate by the $\rm D_1^{M=5}$
{\it Ansatz}, and results of $E_{\rm ex}, E_{\rm ph}$ and $E_{\rm
diag}$ by the $\rm D_1^{M=5}$ {\it Ansatz} are in perfect agreement
with those obtained with the $\rm D_2^{M=32}$ {\it Ansatz}, which
indicates the robustness of the polaron dynamics based on the
multiple Davydov trial states when the multiplicity $M$ is
sufficiently large.

A comprehensive test of the validity for our new trial states consisting of
the multiple Davydov {\it Ans\"atze} is performed for various
parameters sets $(J,W,g, \phi)$. In Fig.~\ref{g=0.1}(a), the
relative deviation $\sigma$, given by Eq.~(\ref{relative_error}), is
displayed as a function of $1/M$, for the diagonal coupling case of
$J=0.1, W=0.5, g=0.1$ and $\phi = 0$. As $M$ increases, the relative
error $\sigma$ monotonically deceases, and the value $\sigma =0.067$
obtained at $1/M=0.2$ is very small, which indicates the length of
the deviation vector $\vec{\delta}(t)$, as defined in
Eqs.~(\ref{deviation_1}), is negligibly small with respect to those
of the vectors $\vec{\chi}(t)$ and $\vec{\gamma}(t)$. Moreover, the
result that the ${\rm D}_1^{M=5}$ {\it Ansatz} is comparable with
$\sigma=0.033$ obtained with the ${\rm D}_2^{M=32}$ {\it Ansatz}
demonstrates the accuracy of the multiple Davydov trial states when
$M$ is sufficiently large.

\begin{figure}[tbp]
\centering
\vspace{1.5\baselineskip}
\includegraphics[scale=0.35]{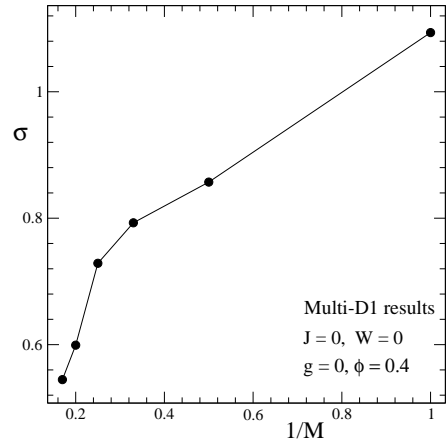}
\caption{The relative deviation $\sigma$ from the multi-${\rm D}_1$ {\it Ansatz} is displayed as a function of $1/M$ for the the off-diagonal coupling case with
the strength $\phi=0.4$, and other parameters $J=W=g=0$ are set.}
\label{phi_diff_M}
\end{figure}

In Fig.~\ref{g=0.1}(b), the relative deviation $\sigma$ is displayed
as a function of the transfer integral $J$ with circles and
triangles corresponding to $M=1$ and $4$ of the multi-$\rm D_1$ {\it
Ans\"atze}, respectively. Other parameters used in the simulation
are $g=0.1, W=0.5$ and $\phi = 0$. An obvious reduction in the
relative error $\sigma$ has been found when the multiplicity $M$ is
increased for the entire $J$ regime. Similarly, the relative error
$\sigma$ against the diagonal coupling strength $g$ is displayed in
Fig.~\ref{g=0.1}(c) for $M=1$ and $4$, respectively. The relative
error $\sigma$ is obviously reduced for the multiplicity $M=4$ in
comparison with that of $M=1$ when $g<0.3$. However, these two
curves overlap for $g>0.3$ as the exciton is self-trapped in one of
the sites. The above results indicate that the multiple Davydov
trial states will significantly improve the accuracy of the
delocalized state, while in the localized state the single $\rm D_1$
{\it Ansatz} is sufficient. In addition, the multiple Davydov trial
states in the off-diagonal coupling case are also investigated
with the nonzero value of $\phi$. Taking the set of parameters
$\phi=0.4$ and $g=J=W=0$ as an example, the relative error $\sigma$
is displayed as a function of $1/M$ in Fig.~\ref{phi_diff_M}. As $M$
increases, the relative error $\sigma$ decreases, similar to the
diagonal coupling case as shown in Fig.~\ref{g=0.1}(a), although the
value of $\sigma$ for $M=6$ ($\sigma=0.54$) remains somewhat large.
For off-diagonal coupling, considerable improvements in accuracy can be achieved
by utilizing multi-$\mathrm{D}_2$ with the increase of multiplicity M
(see discussions in Ref. \cite{zhou_15}).

\subsection{Exciton probabilities and phonon displacements }

Dynamical properties of the Holstein polaron, including the exciton
probabilities and phonon displacements, are investigated by using
the multiple Davydov trial states, in comparison with those obtained
with the single Davydov {\it Ansatz} and the numerically exact HEOM
method \cite{Tanimura1,Tanimura2,Tanimura3,Ishizaki}(see Appendix B). Figure~\ref{HEOM_1} illustrates the time
evolution of the exciton probability $P_{\rm ex}(t, n)$ for the case
of $J=0.5, W=0.5,g=0.1$ and $\phi=0$. For simplicity, a small ring
with $N=10$ sites is used in the simulations. As depicted in
Figs.~\ref{HEOM_1}(a) and \ref{HEOM_1}(b), distinguishable deviation
in $P_{\rm ex}(t, n)$ can be found between the variational results
from the $\rm D^{M=1}_1$ and $\rm D^{M=8}_1$ {\it Ans\"atze}.
Interestingly, the exciton probability $P_{\rm ex}(t, n)$ obtained
from the HEOM method in Fig.~\ref{HEOM_1}(c) almost overlaps with
that in Fig.~\ref{HEOM_1}(b) by the $\rm D_1^{M=8}$ {\it Ansatz}.
Furthermore, the exciton probability difference between the
variational method and the HEOM method, $\Delta P_{\rm ex}(t, n)$,
as depicted in Fig.~\ref{HEOM_1}(d), is two order of magnitude
smaller than the value of $P_{\rm ex}(t, n)$. It indicates that the
variational dynamics of the Holstein polaron can be numerically
exact if the multiplicity $M$ of the $\rm D_1$ {\it Ansatz} is
sufficiently large. In Fig.~\ref{HEOM_2}, the exciton probabilities
$P_{\rm ex}(t, n)$ at the site $n=5$ and $10$ are plotted in the top
and the bottom panels with the solid line, the dashed line and the
circles, corresponding to the variational results obtained with the
single $\rm D_1$ and $\rm D_1^{M=8}$ {\it Ans\"atze} and the HEOM
results, respectively. The near overlap of the dashed line and the
circles further reconfirms the validity of the multi-$\rm D_1$
{\it Ansatz}.

\begin{figure}[tbp]
\centering
\includegraphics[scale=0.4]{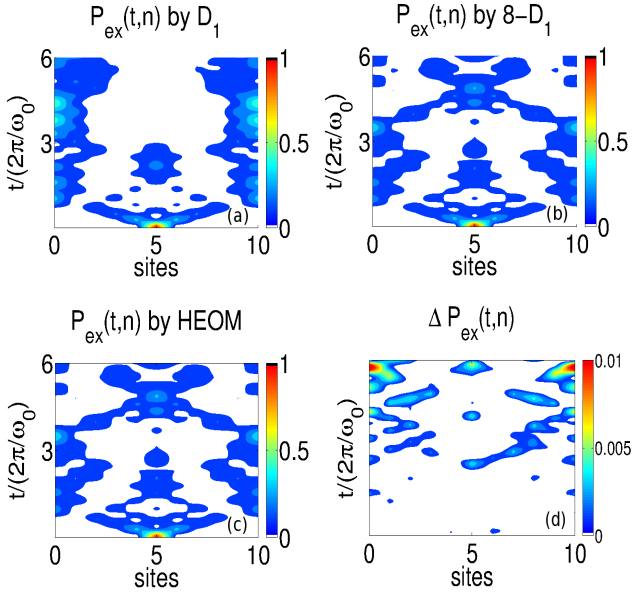}
\caption{ Time evolution of the exciton probability $P_{\rm ex}(t, n)$ for the case of $J=0.5, W=0.5,g=0.1$ and $\phi=0$ is displayed in (a), (b) and (c), corresponding to the results obtained with the single $\rm D_1$ {\it Ansatz}, the $\rm D_1^{M=8}$ {\it Ansatz} and the HEOM method, respectively. The
difference $\Delta P_{\rm ex}(t, n)$ between the HEOM and the $\rm D_1^{M=8}$ variational method is also displayed in (d). For simplify, we set the size of the molecular ring $N=10$ in simulations.
}
\label{HEOM_1}
\end{figure}

\begin{figure}[bp]
\centering
\vspace{1.0\baselineskip}
\includegraphics[scale=0.4]{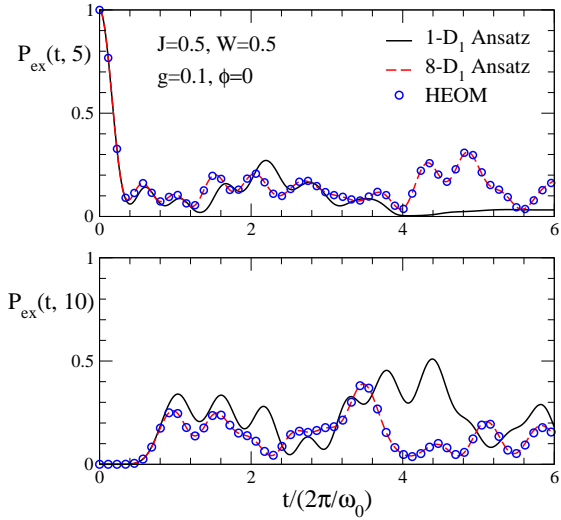}
\caption{ Time evolution of the exciton probability $P_{\rm ex}(t, n)$ with $n=5$ and $10$ are displayed in the top and bottom panel for the case of
$J=0.5, W=0.5,g=0.1$ and $\phi=0$. In each panel, the solid line, dashed line and circles correspond to the variational results obtained with the single $\rm D_1$
and $\rm D_1^{M=8}$ {\it Ans\"atze} and the HEOM results, respectively. The size of the molecular ring is set to $N=10$.
}
\label{HEOM_2}
\end{figure}

\begin{figure}[tbp]
\centering
\includegraphics[scale=0.4]{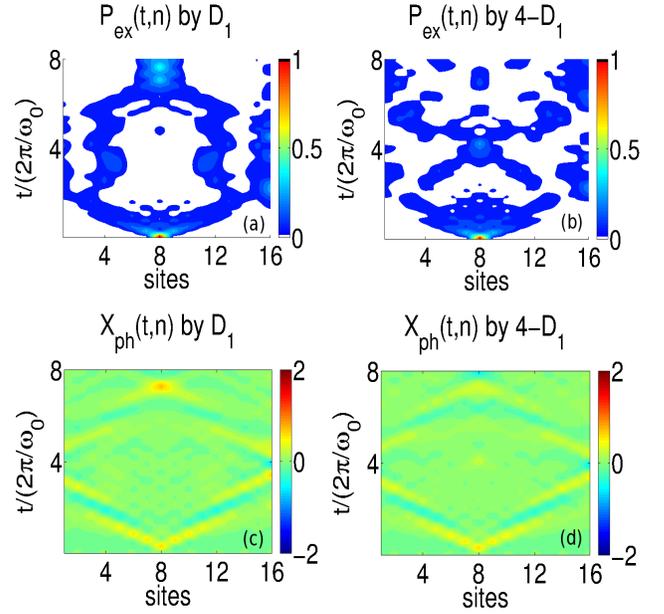}
\caption{ Time evolution of the exciton probability $P_{\rm ex}(t, n)$ and the phonon displacement $X_{\rm ph}(t, n)$  obtained with the single $\rm D_1$ {\it Ansatz} (left panel)
and the ${\rm D}_1^{M=4}$ {\it Ansatz} (right panel) are displayed in (a)-(d) for the case of $W=0.5, g=0.1, J=0.5$ and $\phi=0$.  }
\label{wave_1}
\end{figure}

Displayed in Figs.~\ref{wave_1}(a) and \ref{wave_1}(c) are the exciton probability $P_{\rm ex}(t, n)$ and the phonon displacement $X_{\rm ph}(t, n)$ obtained with the single $\rm D_1$ {\it Ansatz}, respectively, for the case of $W=0.5, g=0.1, J=0.5$ and $\phi=0$. For comparsion, corresponding results of $P_{\rm ex}(t, n)$ and $X_{\rm ph}(t, n)$ obtained by the multi-$\rm D_1$ {\it Ansatz} with $M=4$ are presented in Figs.~\ref{wave_1}(b) and \ref{wave_1}(d), respectively. Quite obvious difference is found in the excitonic behavior for the two cases when $t/(2\pi/\omega_0)>3$. To be specific, the exciton probability calculated by the single $\rm D_1$ {\it Ansatz} staggers around two sites in the ring before being eventually trapped near site 8 accompanied by a thickened phonon cloud [cf.~Fig.~\ref{wave_1}(c)], while that obtained by the multi-$\rm D_1$ {\it Ansatz} with $M=4$ continues to propagate in two opposite directions. The former behavior is apparently an artifact as the combination of $J=0.5$ and $g=0.1$ places the system firmly in the large polaron regime, incompatible with any form of self-trapping at long times. This shows that the single $\rm D_1$ {\it Ansatz} is too simplistic to capture accurate polaron dynamics at long times, especially in the weak coupling regime.

\begin{figure}[tbp]
\centering
\includegraphics[scale=0.4]{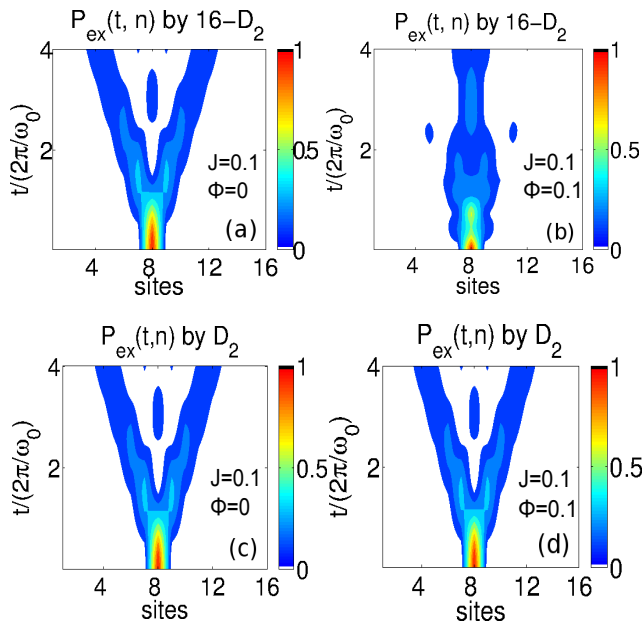}
\caption{ Time evolution of the exciton probability $P_{\rm ex}(t, n)$ is displayed for the case of $J=0.1$ and $\phi=0$ in the left panel
and the case of $J=0.1$ and $\phi=0.1$ in the right panel. Other parameters used are $g=0, W=0.5$ and $N=16$ for both cases.
Two different trial states, the ${\rm D}_2^{M=16}$ and ${\rm D}_2^{M=1}$ {\it Ans\"atze}, are used in the subfigures
(a)-(b) and (c)-(d), respectively. }
\label{wave_2}
\end{figure}

Next, we investigate the improvement on the polaron dynamics by
the multi-$\rm D_2$ trial state. The exciton probability $P_{\rm
ex}(t, n)$ calculated by the multi-$\rm D_2$ {\it Ansatz} with
$M=16$ for two different sets of the parameters, ($J=0.1, g=0,
\phi=0, W=0.5$) and ($J=0.1, g=0, \phi=0.1, W=0.5$), are displayed
in Figs.~\ref{wave_2}(a) and \ref{wave_2}(b), respectively.
Corresponding $P_{\rm ex}(t, n)$ obtained by the single $\rm D_2$
{\it Ansatz} with the same two sets of parameters are shown in
Figs.~\ref{wave_2}(c) and \ref{wave_2}(d), which reveals a similar
pattern of the exciton motion with the same speed of the exciton
packet, $v=\omega_0/2\pi$, despite the jump of the off diagonal
coupling strength from $0$ to $0.1$. In contrast, the exciton
probability obtained with the multi-$\rm D_2$ {\it Ansatz} shows
localization signatures at the off-diagonal coupling strength
$\phi=0.1$, which is absent if $\phi=0$. It indicates that the
combined effect of the transfer integral and the off-diagonal
coupling will confine the exciton to the sites of the initial
creation, despite that acting alone, either the transfer integral or
the off-diagonal coupling may propagate the exciton wave packets.
This phenomenon can be better understood after analyzing the energy
band near the zone center where a discrete self-trapping transition
occurs \cite{zhao_97}. Our calculations show that effective mass in
the case of $\phi=0.1$ is larger than that of $\phi=0$, resulting in
the polaron becoming less mobile. It demonstrates again that the
polaron dynamics obtained with the multiple Davydov trial states is
more accurate than that by the single Davydov trial state.

\subsection{Absorption spectra }

\begin{figure}[tbp]
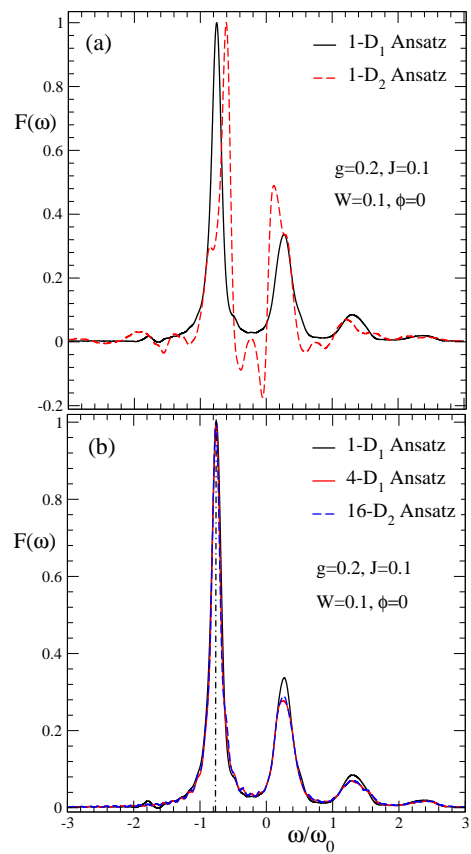

\centering
\includegraphics*[scale=0.35]{figure_10a.eps} \\
\vspace{-1.5\baselineskip}
\includegraphics*[scale=0.35]{figure_10b.eps}
\caption[FIG]{Linear absorption spectra  $F(\omega)$ of a $16$-sites, one-dimensional ring of a coupled exciton-phonon system are displayed in (a)
for the single $\rm D_1$ and $\rm D_2$ {\it Ans\"atze} and in (b) for the single $\rm D_1$, ${\rm D}_1^{M=4}$  and ${\rm D}_2^{M=16}$ {\it Ans\"atze}.
The set of parameters $J=0.1, g=0.2, W=0.1$ and $\phi=0$ are used. A rescaled factor is adopted to normalize the spectral maxima to facilitate comparisons.
The vertical dash-dotted line indicates the location of the zero-phonon line $\omega_{\rm m}/\omega_0=-0.75(1)$.
}
\label{absorption spectra}
\end{figure}

In this subsection we employ the
multiple Davydov trial states to study the linear absorption spectra
$F(\omega)$ defined
in Eq.~(\ref{linear_function}). To facilitate comparisons, spectral
maxima are normalized to unity, and a damping factor of $0.08
\omega_0$ is used \cite{luo_10,sun_10}. In Fig.~\ref{absorption
spectra}, the linear absorption spectra $F(\omega)$ of a $16$-site
ring is displayed for the case of $g=0.2, J=0.1, W=0.1$ and
$\phi=0$. In the subfigure (a), we compare results obtained by the
single $\rm D_1$ (solid) and single $\rm D_2$ (dashed) {\it
Ans\"atze}. Large differences are found between these two curves,
and negative values in the spectrum of the single $\rm D_2$ {\it
Ansatz} point to its apparent invalidity. The multiple $\rm D_1$
trial states are capable to correct such inaccuracies in its
single-$\rm D_1$ counterpart, as demonstrated in the subfigure (b)
for a multiplicity of $4$. Similar corrections are also afforded by
a multi-$\rm D_2$ {\it Ansatz} with a multiplicity of $16$, as shown
in the same panel. Moreover, the position of the zero-phonon line,
denoted by $\omega_{\rm m}$ with respect to $\omega_0$, is marked by
the vertical dash-dotted line at $-0.75(1)$.

\begin{figure}[tbp]
\centering
\includegraphics[scale=.35]{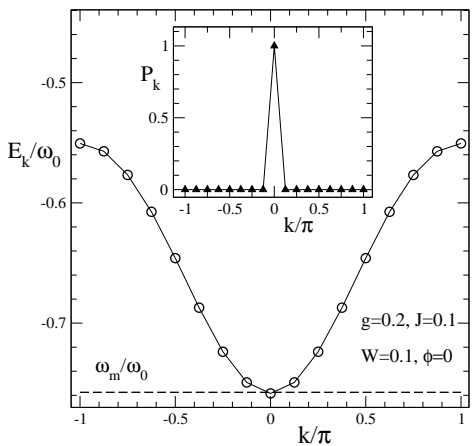}
\caption{Polaron energy bands $E_k/\omega_0$ are calculated
variationally using the Delocalized $\rm D_1$ {\it Ansatz} (solid
line) and the Toyozawa {\it Ansatz} (open circles) for the case of
$g=0.2, J=0.1, W=0.1$ and $\phi=0$. The position of zero-phonon line
$\omega_{\rm m}/\omega_0$ is marked by the dashed line, consistent
with the values of $E_{k=0}/\omega_0$.  A lattice of $N=16$ sites is
used in calculations. In the inset, the transition moment $P_k$ is
plotted as a function of the crystal momentum $k$. }
\label{compare_sd}
\end{figure}

The zero-phonon line can be also determined by the ground-state
polaron energy band $E_{k}$, where $k$ is the crystal momentum. In
order to identify the relationship, the transition moment $P_k$
quantifying the transition probability between the vacuum state and
the exciton state is introduced as $P_k= \left._{\rm ex}\langle 0
|_{\rm ph}\langle 0| \hat{P}^{\dagger}|\Psi_k\rangle\right.$, where
$\hat{P}=\mu\sum_n(\hat{a}_n^{\dagger}|n\rangle_{\rm ex~ex}\langle
n|+|n\rangle_{\rm ex~ex}\langle n|\hat{a}_n)$ is the polarization
operator, and $\Psi_{k}$ is the ground-state trial wave function
with the crystal momentum $k$. By employing the variational method
with the Toyozawa and Delocalized $\rm D_1$ {\it Ans\"atze} (details
are shown in Appendix C), the ground-state wave function $\Psi_k$ can be
obtained, and corresponding polaron energy
band $E_k=\langle \Psi_k|\hat{H}|\Psi_k\rangle$ calculated.
Variations carried out for different $k$ values are independent of
each other, and the set of $E_k$ constitutes a variational estimate
(an upper bound) for the polaron energy band. In
Fig.~\ref{compare_sd}, polaron energy bands $E_k/\omega_0$
calculated variationally for the case of $g=0.2, ~\phi = 0, ~J=0.1$
and $W=0.1$, are plotted as a function of the crystal momentum
$k/\pi$ with the solid and open circles, corresponding to the
Delocalized $\rm D_1$ and Toyozawa {\it Ans\"atze}, respectively.
For simplicity, we set $\mu = \omega_0 = 1$. Interestingly, the
normalized position of the zero-phonon line, $\omega_{\rm
m}/\omega_0$ in Fig.~\ref{absorption spectra}(b), is consistent with
the value of $E_{k=0}/\omega_0$. It indicates that $\Psi_{k=0}$ is the bright
state responsible for the zero-phonon line, in perfect agreement
with the obtained transition probability $P_k$, which is nonzero
only at the crystal momentum $k=0$ as depicted in the inset.

\begin{figure}[bp]
\centering
\vspace{1.0\baselineskip}
\includegraphics[scale=0.35]{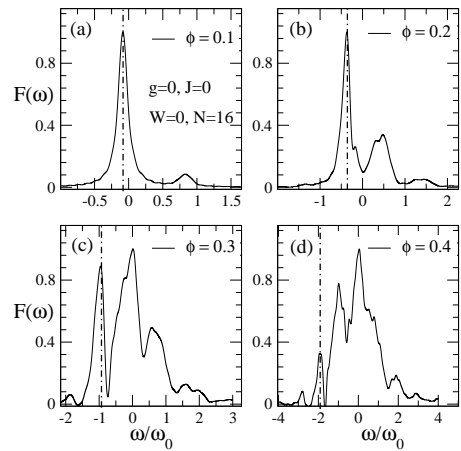}
\caption{ Linear absorption spectra $F(\omega)$ obtained with the
$\rm D_2^{M=16}$ {\it Ansatz} are displayed in (a)-(d) for the
off-diagonal coupling cases with the nonzero coupling strengths
$\phi=0.1, 0.2, 0.3$ and $0.4$, respectively. Other parameters
$g=W=J=0$ and $N=16$ are set. The vertical dash-dotted lines
indicate locations of zero-phonon lines. } \label{spectral_off1}
\end{figure}

\begin{figure}[bp]
\centering
\vspace{1.0\baselineskip}
\includegraphics[scale=.35]{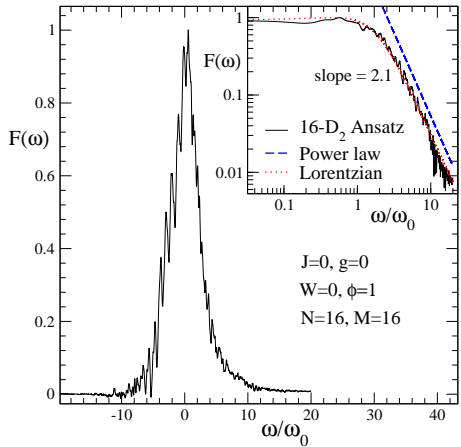}
\caption{ Linear absorption spectrum $F(\omega)$ of the $\rm
D_2^{M=16}$ {\it Ansatz} is displayed for the off-diagonal coupling
case with $\phi=1$. In the inset, the power-law and Lorenz fittings
are given in the log-log scale with the dashed and dotted lines,
respectively. } \label{spectral_off3}
\end{figure}

Moreover, absorption spectra in the presence of
off-diagonal coupling ($\phi \neq 0$) are investigated with the aid of a multi-$\rm
D_2$ {\it Ansatz} with $M=16$ (we set $J=g=W=0$ for simplicity). As
shown in Fig.~\ref{spectral_off1}, with an increase in the
off-diagonal coupling strength $\phi$, phonon sidebands of the
linear absorption spectra become broadened and the intensity of the
zero-phonon line is reduced. Vertical dashed lines shown in the $4$
panels of Fig.~\ref{spectral_off1} denote the positions of the
zero-phonon lines ($\omega_{\rm m}/\omega_0=-0.08, 0.369, -0.956$
and $-1.93$). For strong off-diagonal coupling, such
as the case of $\phi=1$, the linear absorption spectra, shown in
Fig.~\ref{spectral_off3}, behave quite differently from those in
weak off-diagonal coupling cases, such as $\phi=0.1$ and $0.2$
(cf.~Fig.~\ref{spectral_off1}). All of the sharp peaks are smeared
out, and the zero-phonon line almost disappears. In order to better
understand the line shape, we plot the absorption spectrum in a
log-log scale in the inset. A power-law fitting (dashed line) yields a
slope of $2.1(1)$ indicating that the phonon sideband deviates from
the Gaussian line shape. A Lorentzian line-shape function (dotted
line) is then introduced for the fitting, consistent with the
absorption spectrum obtained from the variational method.

\subsection{2D spectra}

\begin{figure}[tbp]
\centering
\includegraphics[scale=.2]{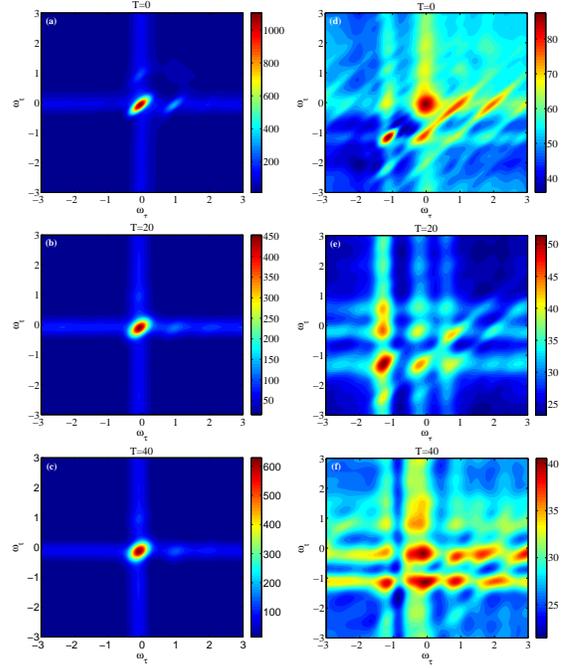}
\caption{2D spectra of the molecular ring for off-diagonal coupling
strengths $\phi=0.1$ (left column) and $\phi=0.4$ (right column).
Upper, middle and lower panels correspond to the population time
$T=0,20,40$, respectively. Other parameters $g=W=J=0$ and $N=10$ are
set.} \label{2D}
\end{figure}

In addition to the linear absorption spectra, fast and accurate
implementation of the multidimensional spectroscopy is possible via
the time-dependent variational method developed here. As an example, we present
in this subsection $2$D
spectra calculated for a molecular ring of $10$ sites using the
multiple ${\rm D}_2$ {\it Ansatz}. For the secondary bath whose
spectral density is defined by Eq.~(\ref{SD}), we adopt the
overdamped Brownian oscillator model with the Drude-Lorentz type
spectral density
\begin{equation}
D(\omega)=2\eta\frac{\gamma\omega}{{\omega}^2+{\gamma}^2}
\end{equation}
The resulting lineshape function [cf.~Eq.~(\ref{LS})] can be
evaluated analytically \cite{Mukamel},
\begin{eqnarray}
g(t)&=&\frac{\eta}{\gamma}\coth\frac{\gamma\beta}{2}[e^{-\gamma{t}}+\gamma{t}-1]-i\frac{\eta}{\gamma}[e^{-\gamma{t}}+\gamma{t}-1]\nonumber \\
&&+\frac{4\eta\gamma}{\beta}\sum_{n=1}^{\infty}\frac{e^{-\nu_nt}+\nu_nt-1}{\nu_n(\nu_n^2-{\gamma}^2)},
\end{eqnarray}
where $\nu_n=2\pi{n}/\beta$ is the Matsubara frequency. In our
calculations, we set $\eta=0.1,\beta=5$ and $\gamma=0.02$.

In Fig. \ref{2D}, $2$D spectra of the $10$-site ring are displayed
for the case of $\phi=0.1$ (left panel) and $\phi=0.4$ (right
panel). For simplicity, we set $J=g=W=0$, and adopt the toy model of
J-aggregates with the tangential (head-to-tail) orientations of the
transition dipoles. We first consider weak off-diagonal coupling.
The $2$D spectra are shown in Figs.~\ref{2D}(a),(b), and (c)
corresponding to the population time $T=0, 20, 40$, respectively. At
$T=0$, the signal exhibits a single peak located at
$(\omega_{\tau},\omega_t)=(-0.08,-0.08)$, which is elongated along
the diagonal line. As the population time increases, the elongation
becomes less pronounced, and the peak appears more rounded. We then
study the case of strong off-diagonal coupling with $\phi=0.4$, as
depicted in the right panel of Fig.~\ref{2D} for several values of
the population time (see Figs.~\ref{2D}(d), (e) and (f) for $T= 0,
20$, and 40, respectively). Overall, it is found that strong exciton
phonon coupling induces a pronounced vibronic multi-peak structure
in the 2D spectra. With increasing population time, the shapes as
well as the strengths for the peaks change, and we also find
population cascades from high to low energy regions with lower
$\omega_t$ for larger values of $T$, as demonstrated in
Figs.~\ref{2D}(d), (e), and (f).

\section{Conclusions}
In this work, we have studied the dynamical properties of the
Holstein polaron in a one-dimensional molecular ring using the
Dirac-Frenkel time-dependent variational principle and an extended
form of the Davydov trial states, also known as the ``multi-D$_1$
{\it Ansatz}" (``multi-D$_2$ {\it Ansatz}"), which is a linear
combination of the usual (single) Davydov $\rm D_1$ ($\rm D_2$)
trial states. For both diagonal and off-diagonal exciton-phonon
coupling, the relative error quantifying how closely the trial state
follows the Schr\"odinger equation is found to decrease with the
multiplicity $M$, reflecting the improvement in accuracy of the
multiple Davydov trial states. Moreover, exciton probabilities
calculated by the multiple Davydov trial states are obtained, in
perfect agreement with those from a numerically exact approach
employing the hierarchy equations of motion, demonstrating the great
promise the multiple Davydov trial states hold as an efficient,
robust description of dynamics of the complex quantum systems.

An abnormal self-trapping phenomenon is uncovered in the dynamical
behavior of polaron with the increase of the off-diagonal coupling.
Besides, the optical spectrum is also studied as a sensitive
indicator of the accuracy of the variational polaron dynamics. Among
our findings, linear absorption spectra from the multi-$\rm D_1$
{\it Ansatz} with a multiplicity of $4$ can be reproduced by the
multi-$\rm D_2$ {\it Ansatz} with a multiplicity of $16$, and the
positions of the zero-phonon lines are in good agreement with
ground-state energy bands calculated by the Toyozawa and the
Delocalized D$_1$ {\it Ans\"atze} in the weak coupling (transfer
integral) regime. Moreover, for the first time, $2$D spectra have been
calculated for systems with off-diagonal exciton-phonon coupling by
employing the multiple ${\rm D}_2$ {\it Ansatz} to compute the
nonlinear response function, testifying to the great potential
of the multiple ${\rm D}_2$ {\it Ansatz}
for fast, accurate implementation of multidimensional
spectroscopy. It is also found that the signal exhibits a
single peak for weak off-diagonal coupling, while a vibronic
multi-peak structure appears for strong off-diagonal coupling.

\section*{Acknowledgments}

The authors thank Vladimir Chernyak for insightful discussion and Jiangfeng Zhu for help with numerics.
Support from the Singapore National Research Foundation through the Competitive Research Programme (CRP) under Project No.~NRF-CRP5-2009-04
is gratefully acknowledged. One of us (N.J.~Zhou) is also supported in part by National Natural Science Foundation of China under Grant No.~$11205043$. K. W. Sun is supported in part by National Natural Science Foundation of China under Grant No.~$ 11404084$ and $11574052$.

\appendix

\section{The Multi-${\rm D}_1$ trial state}

The individual energy terms can be respectively calculated as follows
\begin{eqnarray}
\label{A1}
&&\left\langle {\rm D}^M_{1}\left(t\right)\right|H_{ex}\left|{\rm D}^M_{1}\left(t\right)\right\rangle \nonumber \\
&& =  -J\sum_{i,j}^{M}\sum_{n}\psi_{j,n}^{*}\left(t\right)\left[\psi_{i,n+1}\left(t\right)S_{j,n;i,n+1}\right. \nonumber \\
&&\left. +\psi_{i,n-1}\left(t\right)S_{j,n;i,n-1}\right], \nonumber \\ \\
&&\left\langle {\rm D}^M_{1}\left(t\right)\right|H_{ph}\left|{\rm D}^M_{1}\left(t\right)\right\rangle  \nonumber \\
&& =  \sum_{i,j}^{M}\sum_{n}\psi_{jn}^{\ast}\psi_{in}\sum_{q}\omega_{q}\lambda_{jnq}^{\ast}\lambda_{inq}S_{ji}, \nonumber \\ \\
&&\left\langle {\rm D}^M_{1}\left(t\right)\right|H_{ex-ph}^{diag}\left|{\rm D}^M_{1}\left(t\right)\right\rangle \nonumber \\
&& =  -g\sum_{i,j}^{M}\sum_{n}\psi_{jn}^{\ast}\psi_{in}\sum_{q}\omega_{q}\left(e^{iqn}\lambda_{inq}+e^{-iqn}\lambda_{jnq}^{\ast}\right)S_{ji} , \nonumber \\ \\
&&\left\langle {\rm D}^M_{1}\left(t\right)\right|H_{ex-ph}^{o.d.}\left|{\rm D}^M_{1}\left(t\right)\right\rangle  \nonumber \\
&& =  \frac{1}{2}\phi\sum_{n,q}\sum_{i,j}^{M}\omega_{q}S_{j,n;i,n+1}\psi_{j,n}^{*}\left(t\right)\psi_{i,n+1}\left(t\right) \nonumber \\
&& \left[e^{iqn}\left(e^{iq}-1\right)\lambda_{i,n+1,q}\left(t\right)+e^{-iqn}\left(e^{-iq}-1\right)\lambda_{jnq}^{*}\left(t\right)\right] \nonumber \\
&&  +  \frac{1}{2}\phi\sum_{n,q}\sum_{i,j}^{M}\omega_{q}S_{j,n;i,n-1}\psi_{j,n}^{*}\left(t\right)\psi_{i,n-1}\left(t\right)\nonumber \\
&& \left[e^{iqn}\left(1-e^{-iq}\right)\lambda_{i,n-1,q}\left(t\right)+e^{-iqn}\left(1-e^{iq}\right)\lambda_{jnq}^{*}\right], \nonumber \\
\end{eqnarray}
where the Debye-Waller factor is formulated as
\begin{eqnarray}
S_{ij} & = & \left\langle \lambda_{i}\vert\lambda_{j}\right\rangle, \nonumber \\
S_{j,n;i,n+1} & = & \left\langle \lambda_{j,n}\vert\lambda_{i,n+1}\right\rangle,
\label{A2}
\end{eqnarray}
The Dirac-Frenkel variational principle leads to equations of motion:
\begin{eqnarray}
&&-i\sum_{i}^{M}\dot{\psi}_{in}S_{ki}\nonumber \\
&&-\frac{i}{2}\sum_{i}^{M}\psi_{in}\sum_{q}\left(2\lambda_{knq}^{\ast}\dot{\lambda}_{inq}-\dot{\lambda}_{inq}
\lambda_{inq}^{\ast}-\lambda_{inq}\dot{\lambda}_{inq}^{\ast}\right)S_{k,i}\nonumber \\
&&= J\sum_{i}^{M}\left(\psi_{i,n+1}S_{k,n;i,n+1}+\psi_{i,n-1}S_{k,n;i,n-1}\right) \nonumber \\ &&-\sum_{i}^{M}\psi_{in}\sum_{q}\omega_{q}\lambda_{knq}^{\ast}\lambda_{inq}S_{ki}\nonumber \\
&& + g\sum_{i}^{M}\psi_{in}\sum_{q}\omega_{q}\left(e^{iqn}\lambda_{inq}+e^{-iqn}\lambda_{knq}^{\ast}\right)S_{ki} \nonumber \\
&& -\frac{1}{2}\phi\sum_{i}^{M}\sum_{q}\omega_{q}\psi_{i,n+1}\left(t\right)\left[e^{iqn}\left(e^{iq}-1\right)\lambda_{i,n+1,q}\left(t\right)\right.\nonumber \\
&&\left. +e^{-iqn}\left(e^{-iq}-1\right)\lambda_{knq}^{*}\left(t\right)\right]S_{k,n;i,n+1}\nonumber \\
&&-\frac{1}{2}\phi\sum_{i}^{M}\sum_{q}\omega_{q}\psi_{i,n-1}\left(t\right)\left[e^{iqn}\left(1-e^{-iq}\right)\lambda_{i,n-1,q}\left(t\right)\right. \nonumber \\
&&\left.+e^{-iqn}\left(1-e^{iq}\right)\lambda_{knq}^{*}\right]S_{k,n;i,n-1}; \label{A3} \\
&& -i\sum_{i}^{M}\psi_{kn}^{\ast}\dot{\psi}_{in}\lambda_{inq}S_{ki} \nonumber \\
&& -i\sum_{i}^{M}\psi_{kn}^{\ast}\psi_{in}\dot{\lambda}_{inq}S_{ki} \nonumber \\
&& -\frac{i}{2}\sum_{i}^{M}\psi_{kn}^{\ast}\psi_{in}\lambda_{inq}S_{k,i}\nonumber \\
&& \sum_{p}\left(2\lambda_{knp}^{\ast}\dot{\lambda}_{inp}-\dot{\lambda}_{inp}
\lambda_{inp}^{\ast}-\lambda_{inp}\dot{\lambda}_{inp}^{\ast}\right) \nonumber \\
&& = J\sum_{i}^{M}\psi_{k,n}^{*}\left(\psi_{i,n+1}\lambda_{i,n+1,q}S_{k,n;i,n+1}\right.\nonumber \\
&& \left.+\psi_{i,n-1}\lambda_{i,n-1,q}S_{k,n;i,n-1}\right)\nonumber \\
&& -\sum_{i}^{M}\psi_{kn}^{\ast}\psi_{in}\lambda_{inq}S_{ki}\left(\omega_{q}+\sum_{p}\omega_{p}\lambda_{knp}^{\ast}\lambda_{inp}\right)\nonumber \\
&& + g\sum_{i}^{M}\psi_{kn}^{\ast}\psi_{in}\omega_{q}e^{-iqn}S_{ki} \nonumber \\
&& +g\sum_{i}^{M}\psi_{kn}^{\ast}\psi_{in}\lambda_{inq}\sum_{p}\omega_{p}\left(e^{ipn}\lambda_{inp}+e^{-ipn}\lambda_{knp}^{\ast}\right)S_{k,i} \nonumber \\
&& -\frac{1}{2}\phi\sum_{i}^{M}\omega_{q}\psi_{kn}^{*}\left[\psi_{i,n+1}e^{-iqn}\left(e^{-iq}-1\right)S_{k,n;i,n+1}\right.\nonumber \\
&&\left.+\psi_{i,n-1}e^{-iqn}\left(1-e^{iq}\right)S_{k,n;i,n-1}\right]\nonumber \\
&& -\frac{1}{2}\phi\sum_{p}\sum_{i}^{M}\omega_{p}\psi_{k,n}^{*}\psi_{i,n+1}\left[e^{ipn}\left(e^{ip}-1\right)\lambda_{i,n+1,p}\right.\nonumber \\
&&\left.+e^{-ipn}\left(e^{-ip}-1\right)\lambda_{knp}^{*}\right]\lambda_{i,n+1,q}S_{k,n;i,n+1}\nonumber \\
&& -\frac{1}{2}\phi\sum_{p}\sum_{i}^{M}\omega_{p}\psi_{k,n}^{*}\psi_{i,n-1}\left[e^{ipn}\left(1-e^{-ip}\right)\lambda_{i,n-1,p}\right.\nonumber \\
&&\left.+e^{-ipn}\left(1-e^{ip}\right)\lambda_{knp}^{*}\right]\lambda_{i,n-1,q}S_{k,n;i,n-1}. \label{A4}
\end{eqnarray}

\section{Hierarchy equation of motion}

For the Holstein model [Eq.~(\ref{Holstein})], let us denote $|n\rangle=\hat{a}_n^{\dagger}|0\rangle_{\mathrm{ex}}$ where $|0\rangle_{\mathrm{ex}}$ stands for the exciton vacuum. Then the reduced density matrix element for the exciton system is expressed in the path integral form with the factorized initial condition as \cite{LP_HEOM}
\begin{eqnarray}\label{Rho}
&& \rho(n,n';t) = \\
&& \int\mathcal{D}n\int\mathcal{D}n'\rho(n_0,n_0^{'};t_0)\times{e^{iS[n;t]}}F(n,n';t)e^{-iS[n';t]}, \nonumber
\end{eqnarray}
where $S[n]$ is an action of the exciton system, and $F[n,n']$ is the Feynman-Vernon influence functional
\begin{eqnarray}\label{IF}
&&F[n,n']\nonumber \\
&&=\exp(-\sum_q\omega_q^2\int_{t_0}^tds\int_{t_0}^sds'{V_q^{*}}^{\times}(s)\nonumber \\
&& \times[{V_q}^{\times}(s')\coth(\beta\omega_q/2)\cos(\omega_q(s-s'))\nonumber \\
&& -i{V_q}^{\circ}(s')\sin(\omega_q(s-s'))]).
\end{eqnarray}
In the above equation, $\beta$ is the inverse of temperature ($\beta=1/k_BT$), and the abbreviations
\begin{equation}
{V_q}^{\times}=V_q(n)-V_q(n') \quad \quad {V_q}^{\circ}=V_q(n)+V_q(n'),
\end{equation}
 are introduced with $\hat{V}_q^{\dagger}=g\sum_n\hat{a}_n^{\dagger}\hat{a}_ne^{iqn}$.

Equation (\ref{IF}) can be rewritten as
\begin{eqnarray}\label{IFS}
&&F[n,n']\nonumber \\
&&=\exp(-\sum_q\omega_q^2\int_{t_0}^tds\int_{t_0}^sds'{V_q^{*}}^{\times}(s)  \\
&& \times[\frac{e^{i\omega_q(s-s')}}{2}({V_q}^{\times}(s')\coth{(\beta\omega_q/2)}-{V_q}^{\circ}(s'))
\nonumber \\
&&+\frac{e^{-i\omega_q(s-s')}}{2}({V_q}^{\times}(s')\coth(\beta\omega_q/2)+{V_q}^{\circ}(s'))]), \nonumber
\end{eqnarray}
Taking derivative of Eq.(\ref{Rho}), one has
\begin{eqnarray}\label{RhoS}
&&\frac{\partial}{\partial{t}}\rho(n,n';t)\nonumber \\
&&=-i\mathcal{L}\rho(n,n';t)  \\
&&-\sum_q\omega_q^2{V_q^{*}}^{\times}(t)\int\mathcal{D}n\int\mathcal{D}n'\rho(n_0,n_0^{'};t_0)\nonumber \\
&&\int_0^{t}ds'[\frac{e^{i\omega_q(t-s')}}{2}({V_q}^{\times}(s')\coth(\beta\omega_q/2)-{V_q}^{\circ}(s'))\nonumber \\
&&+\frac{e^{-i\omega_q(t-s')}}{2}({V_q}^{\times}(s')\coth(\beta\omega_q/2)+{V_q}^{\circ}(s'))]\nonumber \\
&&\times{e^{iS[n,t]}}F(n,n';t)e^{-iS[n',t]}, \nonumber
\end{eqnarray}
If we use the following super-operator
\begin{eqnarray}
\hat{\Phi}_q(t)&=&\omega_q^2{V_q^{*}}^{\times}(t)/2, \nonumber \\
\hat{\Theta}_{q\pm}(t)&=&{V_q}^{\times}(t)\coth(\beta\omega_q/2)\mp{V_q}^{\circ}(t),
\end{eqnarray}
Eqs.(\ref{IFS}) and (\ref{RhoS}) then can be simplified as
\begin{eqnarray}
&&F[n,n']\nonumber\\
&&=\exp(-\sum_q\int_{t_0}^tds\int_{t_0}^sds'\Phi_q(s)\times [e^{i\omega_q(s-s')}\Theta_{q+}(s')\nonumber \\
&&+e^{-i\omega_q(s-s')}\Theta_{q-}(s')]),
\end{eqnarray}
\begin{eqnarray}
&&\frac{\partial}{\partial{t}}\rho(n,n';t)\nonumber \\
&&= -i\mathcal{L}\rho(n,n';t) \nonumber \\ &&-\sum_q\Phi_q(t)\int\mathcal{D}n\int\mathcal{D}n'\rho(n_0,n_0^{'};t_0)  \\
&& \int_0^tds'[e^{i\omega_q(t-s')}\Theta_{q+}(s') \nonumber \\
&&+e^{-i\omega_q(t-s')}\Theta_{q-}(s')]\times{e^{iS[n,t]}}F(n,n';t)e^{-iS[n',t]}, \nonumber
\end{eqnarray}
In order to derive the equations of motion, we introduce the auxiliary operator $\rho_{m_{1\pm},m_{2\pm},\cdots,m_{N\pm}}(n,n';t)$ by its matrix element as
\begin{eqnarray}
&&\rho_{m_{1\pm},m_{2\pm},\cdots,m_{N\pm}}(n,n';t)=   \\
&&\int\mathcal{D}n\int\mathcal{D}n'\rho(n_0,n_0^{'};t_0)\prod_{q=1}^{N}
(\int_{t_0}^tdse^{i\omega_q(t-s)}\Theta_{q+}(s))^{m_{q+}}\nonumber \\
&&(\int_{t_0}^{t}dse^{-i\omega_q(t-s)}\Theta_{q-}(s))^{m_{q-}}\times{e^{iS[n,t]}}F(n,n')e^{-iS[n',t]}, \nonumber
\end{eqnarray}
for non-negative integers $m_{1\pm},m_{2\pm},...,m_{N\pm}$. Note that only $\hat{\rho}_{0......0}(t)=\hat{\rho}(t)$ has a physical meaning and the others are introduced for computational purposes only. Differentiating $\rho_{m_{1\pm},m_{2\pm},...,m_{N\pm}}(n,n';t)$ with respect to $t$, we can obtain the following hierarchy of equations in the operator form
\begin{eqnarray}\label{HEOM}
&&\frac{\partial}{\partial{t}}\hat{\rho}_{m_{1\pm},\cdots,m_{N\pm}}(t)\nonumber \\
&&=-i\mathcal{L}\hat{\rho}_{m_{1\pm},\cdots,m_{N\pm}}(t)\nonumber \\
&&-i\sum_q\omega_q(m_{q-}-m_{q+})\hat{\rho}_{m_{1\pm},\cdots,m_{N\pm}}(t) \nonumber \\
&&-\sum_q\hat{\Phi}_q(\hat{\rho}_{m_{1\pm},\cdots,m_{q+}+1,m_{q-},\cdots,m_{N\pm}}(t) \nonumber \\
&&+ \hat{\rho}_{m_{1\pm},\cdots,m_{q+},m_{q-}+1,\cdots,m_{N\pm}}(t))\nonumber \\
&&+\sum_q(m_{q+}\hat{\Theta}_{q+}\hat{\rho}_{m_{1\pm},\cdots,m_{q+}-1,m_{q-},\cdots,m_{N\pm}}(t)\nonumber \\
&&+ m_{q-}\hat{\Theta}_{q-}\hat{\rho}_{m_{1\pm},\cdots,m_{q+},m_{q-}-1,\cdots,m_{N\pm}}(t)),
\end{eqnarray}

The HEOM consists of an infinite number of equations, but they can be truncated using a number of hierarchy elements. The infinite hierarchy of Eq.(\ref{HEOM}) can be truncated by the terminator as
\begin{eqnarray}
&&\frac{\partial}{\partial{t}}\hat{\rho}_{m_{1\pm},\cdots,m_{N\pm}}(t)  \\
&&=-(i\mathcal{L}+i\sum_q\omega_q(m_{q-}-m_{q+}))\times\hat{\rho}_{m_{1\pm},\cdots,m_{N\pm}}(t). \nonumber
\end{eqnarray}
The total number of hierarchy elements can be evaluated as $L_{\mathrm{tot}}=(N_{\mathrm{trun}}+2N)!/N_{\mathrm{trun}}!(2N)!$, while the total number of termination elements is $L_{\mathrm{term}}=(N_{\mathrm{trun}}+2N-1)!/(2N-1)!N_{\mathrm{trun}}!$, where $N_{\mathrm{trun}}$ is the depth of the hierarchy for $m_{q\pm}(q=1,\cdots,N)$. In practice, we can set the termination elements to zero and thus the number of hierarchy elements for the calculation can be reduced as $L_{\mathrm{calc}}=L_{\mathrm{tot}}-L_{\mathrm{term}}$.

\section{The Delocalized D$_1$ Ansatz and the Toyozawa Ansatz}

Our interest in this work includes the polaron ground-state energy band,
computed as
\begin{eqnarray}
E(\kappa)=\langle \Psi(\kappa) |\hat{H}|\Psi(\kappa) \rangle,
\end{eqnarray}
where$|\Psi(\kappa) \rangle$ is an appropriately normalized, delocalized trial state, and $\hat{H}$ is the system Hamiltonian. The joint crystal
momentum is indicated by the Greek $\kappa$. It should be noted that the crystal momentum operator commutes with the system Hamiltonian, and energy eigenstates are also eigenfunctions of the crystal momentum. Therefore, variations for distinct $\kappa$ are independent. The set of $E(\kappa)$ constitutes a variational estimate (an upper bound) for the polaron energy band. The relaxation iteration technique, viewed as an efficient method for identifying energy minima of a complex variational system, is adopted in this work to obtain numerical solutions to a set of self-consistency equations derived from the
variational principle. To achieve efficient and stable iterations toward the variational ground state, one may take advantage of the continuity of the ground state with respect to small changes in system parameters over most of the phase diagram and may initialize the iteration using a reliable ground state already determined at some nearby points in parameter space. Starting from those limits where exact solutions can be obtained analytically and executing a sequence
of variations along well-chosen paths through the parameter space using solutions from one step to initialize the next, the whole parameter space can be explored.

As the D$_1$ and D$_2$ {\it Ans\"atze} are localized states from the soliton theory, but without considering a form factor of a delocalized state. The polaron state have been analyzed with the delocalized D$_1$ and Toyozawa {\it Ans\"atze}, both of which
are Bloch states with the designated crystal momentum.  The D$_1$ and D$_2$ {\it Ans\"atze} can be delocalized into the delocalized D$_1$ and Toyozawa {\it Ans\"atze}
via a projection operator $\hat P_\kappa$
\begin{eqnarray}
\hat P_\kappa = N^{-1}\underset{n}{\sum }e^{i(\kappa-\hat P) n}=\delta(\kappa-\hat P),
\end{eqnarray}
where
\begin{eqnarray}
\hat P= \underset{k}{\sum }k a_{k}^{\dagger }a_{k}+\underset{q}{\sum }q b_{q}^{\dagger }b_{q},
\end{eqnarray}

The delocalized D$_1$ {\it Ansatz} are then obtained after the delocalization onto the usual $\rm D_1$ {\it Ansatz},
\begin{eqnarray}
|\Psi_1(\kappa)\rangle =|\kappa \rangle \langle \kappa|\kappa \rangle^{-1/2},
\end{eqnarray}

\begin{eqnarray}
 |\kappa \rangle &=& \underset{n}{\sum }e^{i\kappa n}\underset{n1}{\sum }\alpha _{n_1-n}^{\kappa }a_{n_1}^{\dagger } \\
&&\exp [-\underset{n_2}{\sum }(\beta
_{n_1-n,n_2-n}^{\kappa }b_{n_2}^{\dagger }-{\rm H.c.})]|0\rangle, \nonumber
\end{eqnarray}
where $\rm H.c.$ stands for the Hermitian conjugate, $|0\rangle$ is the product
of the exciton and phonon vacuum states, $\alpha _{n_1-n}^{\kappa }$ is the exciton
amplitude, and the phonon displacement $\beta_{n_1-n,n_2-n}^{\kappa }$
depends on $n_1$ and $n_2$, respectively, the sites at which an electronic
excitation and a phonon are generated.

After the delocalization onto the usual ${\rm D}_2$ {\it Ansatz}, the Toyozawa {\it Ansatz} is given by
\begin{eqnarray}
|\Psi_2(\kappa')\rangle =|\kappa' \rangle \langle \kappa'|\kappa' \rangle^{-1/2},
\end{eqnarray}

\begin{eqnarray}
|\kappa' \rangle  & = & \underset{n}{\sum }e^{i\kappa' n}\underset{n1}{\sum }\psi _{n_1-n}^{\kappa' }a_{n_1}^{\dagger } \\
&&\exp [-\underset{n2}{\sum }(\lambda
_{n_2-n}^{\kappa' }b_{n_2}^{\dagger }-{\rm H.c.})]|0\rangle, \nonumber
\end{eqnarray}
where $\psi _{n_1-n}^{\kappa' }$ is the exciton amplitude analogous to $\alpha _{n_1-n}^{\kappa }$
in the delocalized D$_1$ {\it Ansatz}, and $\lambda_{n_2-n}^{\kappa' }$ is the phonon displacement.
Actually, $\lambda_{n_2-n}^{\kappa' }$ is just one column of the phonon displacement matrix $\beta_{n_1-n,n_2-n}^{\kappa }$ in the delocalized D$_1$ {\it Ansatz}.

\section{Simulation of 2D spectra using multiple D$_2$ Ansatze}

In order to describe the population decays and dephasings induced by solvent, we add additional term $H_{B}+H_{SB}$ to the Hamiltonian (\ref{Holstein})
\begin{eqnarray}
\hat{H}&=&\hat{H}_{ex}+\hat{H}_{ph}+\hat{H}_{ex-ph}^{diag}+\hat{H}_{ex-ph}^{o.d.}+\hat{H}_{B}+\hat{H}_{SB}\nonumber \\
&=&\hat{H}_S+\hat{H}_B+\hat{H}_{SB}
\end{eqnarray}
where we have included vibrational modes with significant exciton-phonon coupling into system Hamiltonian, i.e., $\hat{H}_S=\hat{H}_{ex}+\hat{H}_{ph}+\hat{H}_{ex-ph}^{diag}+\hat{H}_{ex-ph}^{o.d.}$, and treated the rest of vibrational modes as a heat bath. We assume a harmonic bath with site-independent and diagonal system bath coupling \cite{SunKeWei1,SunKeWei2,LP_OD}
\begin{eqnarray}
\hat{H}_B&=&\sum_j\hbar\Omega_jc_j^{\dagger}c_j \\
\hat{H}_{SB}&=&\sum_j\sum_{n=1}^{N}\kappa_j\hbar\Omega_j(c_j^{\dagger}+c_j)a_n^{\dagger}a_n
\end{eqnarray}
Here, $c_j(c_j^{\dagger})$ is the annihilation (creation) operator of the $j$th bath mode with frequency $\Omega_j$, and $\kappa_j$ is the corresponding exciton-bath coupling strength. The bath spectral density is specified by
\begin{equation}\label{SD}
D(\omega)=\sum_j\kappa_j^2\Omega_j^2\delta(\omega-\Omega_j)
\end{equation}
It is noted that system-bath Hamiltonian $\hat{H}_{SB}$ commutes with the system Hamiltonian $\hat{H}_S$, and as a result, the nonlinear response function can be represented as a product of the system and bath. Furthermore, by making use of the fact that the system-bath coupling is the same for all excitons, the effect of bath can be taken into account through lineshape factors $F_i$ in the framework of second-order cummulant expansion. Finally, we arrived at the formulas for the nonlinear response function \cite{SunKeWei1}
\begin{eqnarray}
R_1{(t,T,\tau)}&=&F_1(t,T,\tau)\sum_{n,n',n'',n'''}C_{n,n',n'',n'''}\nonumber \\
&&_{ph}\langle{0}|\langle{n}|e^{iH_ST}|n'\rangle\langle{n''}|e^{-H_S(t+T+\tau)}|n'''\rangle|0\rangle_{\rm ph}\nonumber \\
R_2{(t,T,\tau)}&=&F_2(t,T,\tau)\sum_{n,n',n'',n'''}C_{n,n',n'',n'''}\nonumber \\
&&_{ph}\langle{0}|\langle{n}|e^{iH_S(\tau+T)}|n'\rangle\langle{n''}|e^{-H_S(t+T)}|n'''\rangle|0\rangle_{\rm ph}\nonumber \\
R_3{(t,T,\tau)}&=&F_3(t,T,\tau)\sum_{n,n',n'',n'''}C_{n,n',n'',n'''}\nonumber \\
&&_{ph}\langle{0}|\langle{n}|e^{iH_S\tau}|n'\rangle\langle{n''}|e^{-H_St}|n'''\rangle|0\rangle_{\rm ph}\nonumber \\
R_4{(t,T,\tau)}&=&F_4(t,T,\tau)\sum_{n,n',n'',n'''}C_{n,n',n'',n'''}\nonumber \\
&&_{ph}\langle{0}|\langle{n}|e^{-iH_St}|n'\rangle\langle{n''}|e^{-H_S\tau}|n'''\rangle|0\rangle_{\rm ph}
\end{eqnarray}
Here
\begin{equation}
C_{n,n',n'',n'''}=(\bf{e_1}\boldsymbol{\mu_n})(\bf{e_2}\boldsymbol{\mu_{n'}})(\bf{e_3}\boldsymbol{\mu_{n''}})(\bf{e_4}\boldsymbol{\mu_{n'''}})
\end{equation}
are the geometrical factors which must be averaged over orientations of the transition dipole moments $\boldsymbol{\mu_n}$. For simplicity, we can assume all laser fields have the same polarization, then the averaging can be done analytically, leading to
\begin{eqnarray}
C_{n,n',n'',n'''}&=&\frac{1}{15}((\boldsymbol{\mu}_{n}\boldsymbol{\mu}_{n'})(\boldsymbol{\mu}_{n''}\boldsymbol{\mu}_{n'''})\nonumber \\
&&+(\boldsymbol{\mu}_{n}\boldsymbol{\mu}_{n''})(\boldsymbol{\mu}_{n'}\boldsymbol{\mu}_{n'''})+(\boldsymbol{\mu}_{n}\boldsymbol{\mu}_{n'''})(\boldsymbol{\mu}_{n''}\boldsymbol{\mu}_{n'}))\nonumber \\
\end{eqnarray}
The lineshape factors $F_i$ can be easily evaluated as \cite{Mukamel}
\begin{eqnarray}
F_1(t,T,\tau)&=&e^{-g^{*}(t)-g(\tau)-g^{*}(T)+g^{*}(T+t)+g(\tau+T)-g(\tau+T+t)}\nonumber \\
F_2(t,T,\tau)&=&e^{-g^{*}(t)-g^{*}(\tau)+g(T)-g(T+t)-g^{*}(\tau+T)+g^{*}(\tau+T+t)}\nonumber \\
F_3(t,T,\tau)&=&e^{-g(t)-g^{*}(\tau)+g^{*}(T)-g^{*}(T+t)-g^{*}(\tau+T)+g^{*}(\tau+T+t)}\nonumber \\
F_4(t,T,\tau)&=&e^{-g(t)-g(\tau)-g(T)+g(T+t)+g(\tau+T)-g(\tau+T+t)}\nonumber
\\
\end{eqnarray}
where $g(t)$ is the lineshape function
\begin{eqnarray}\label{LS}
g(t)&=&\int_0^{\infty}d\omega\frac{D(\omega)}{\omega^2}\times\nonumber \\
&&\big[\coth\frac{\hbar\omega\beta}{2}(1-\cos{\omega{t}})+i(\sin{\omega{t}}-\omega{t})\big]
\end{eqnarray}

The next crucial step is to approximate the propagator in terms of multiple $D_2$ Ansatze, i.e,
\begin{eqnarray}
&&e^{-iH_st}|n\rangle|0\rangle_{ph}\nonumber \\
&&=\sum_{i}^{M}\sum_n^{N}\psi_{i,n}\hat{a}_n^{\dagger}|0\rangle_{ex}\exp\left\{ \sum_{q}\left[\lambda_{iq}\hat{b}_{q}^{\dagger}-\lambda_{iq}^{\ast}\hat{b}_{q}\right]\right\} \left|0\right\rangle _{\rm ph}\nonumber
\\
\end{eqnarray}
Explicitly, we have final expressions for the nonlinear response function
\begin{eqnarray}
R_1{(t,T,\tau)}&=&F_1(t,T,\tau)\sum_{n,n',n'',n'''}C_{n,n',n'',n'''}\sum_{i,j=1}^{M}\psi_{jn'}^{n*}(T)\nonumber \\
&&\psi_{in''}^{n'''}(\tau+T+t)e^{-\frac{1}{2}\sum_q(|\lambda_{jq}^{n}(T)|^2+|\lambda_{iq}^{n'''}(\tau+T+t)|^2)}\nonumber \\
&&e^{\sum_q\lambda_{jq}^{n*}(T)\lambda_{iq}^{n'''}(\tau+T+t)e^{i\omega_qt}}\nonumber \\
R_2{(t,T,\tau)}&=&F_2(t,T,\tau)\sum_{n,n',n'',n'''}C_{n,n',n'',n'''}\sum_{i,j=1}^{M}\psi_{jn'}^{n*}(T+\tau)\nonumber \\
&&\psi_{in''}^{n'''}(T+t)e^{-\frac{1}{2}\sum_q(|\lambda_{jq}^{n}(T+\tau)|^2+|\lambda_{iq}^{n'''}(T+t)|^2)}\nonumber \\
&&e^{\sum_q\lambda_{jq}^{n*}(T+\tau)\lambda_{iq}^{n'''}(T+t)e^{i\omega_qt}}\nonumber \\
R_3{(t,T,\tau)}&=&F_3(t,T,\tau)\sum_{n,n',n'',n'''}C_{n,n',n'',n'''}\sum_{i,j=1}^{M}\psi_{jn'}^{n*}(\tau)\nonumber \\
&&\psi_{in''}^{n'''}(t)e^{-\frac{1}{2}\sum_q(|\lambda_{jq}^{n}(\tau)|^2+|\lambda_{iq}^{n'''}(t)|^2)}\nonumber \\
&&e^{\sum_q\lambda_{jq}^{n*}(\tau)\lambda_{iq}^{n'''}(t)e^{i\omega_q(t+T)}}\nonumber \\
R_4{(t,T,\tau)}&=&F_4(t,T,\tau)\sum_{n,n',n'',n'''}C_{n,n',n'',n'''}\sum_{i,j=1}^{M}\psi_{jn'}^{n*}(-t)\nonumber \\
&&\psi_{in''}^{n'''}(\tau)e^{-\frac{1}{2}\sum_q(|\lambda_{jq}^{n}(-t)|^2+|\lambda_{iq}^{n'''}(\tau)|^2)}\nonumber \\
&&e^{\sum_q\lambda_{jq}^{n*}(-t)\lambda_{iq}^{n'''}(\tau)e^{-i\omega_qT}}
\end{eqnarray}

\end{document}